%% file: main.tex
\documentclass[journal]{IEEEtran}
\usepackage[binary-units,per-mode=symbol,per-symbol=p]{siunitx}
\usepackage{cite}
\usepackage{adjustbox}
\usepackage[ruled,vlined]{algorithm2e}
\usepackage{amsmath,amssymb,amsfonts}
\usepackage{algorithmic}
\usepackage{graphicx}
\usepackage{textcomp}
\usepackage[colorlinks=false, pdfborder={0 0 0}, breaklinks]{hyperref}
\usepackage[inline]{enumitem}

\usepackage{multirow}
\usepackage{multicol}
\usepackage{subcaption}
\def\BibTeX{{\rm B\kern-.05em{\sc i\kern-.025em b}\kern-.08em
    T\kern-.1667em\lower.7ex\hbox{E}\kern-.125emX}}

\usepackage{amsmath}
\DeclareMathOperator*{\argmax}{arg\,max}

\usepackage{mathtools}

\usepackage{tikz}
\usetikzlibrary{shapes.geometric, arrows}
\usetikzlibrary{fit}
\usetikzlibrary{calc}
\usetikzlibrary{shapes.geometric, arrows}
\tikzstyle{arrow} = [thick,->,>=stealth]

\newcommand*\circled[1]{\tikz[baseline=(char.base)]{%
            \node[shape=circle,fill=blue!20,draw,inner sep=0.5pt] (char) {#1};}}

\usepackage{ctable}
\usepackage{balance}
\newcommand{\etal}{\emph{et al.}\xspace}
\newcommand{\ie}{\emph{i.e.}, }
\newcommand{\eg}{\emph{e.g.}, }

\newcommand{\cf}{\emph{cf.}\xspace}
\newcommand{\HAS}{\emph{HTTP Adaptive Streaming}\xspace}

\newcommand{\VOD}{\emph{Video on Demand }}

\newcommand{\DCT}{\emph{discrete cosine transform }}

\newcommand{\HLS}{\emph{HTTP Live Streaming}\xspace}

\newcommand{\jnd}{\emph{Just Noticeable Difference}\xspace}

\newcommand{\caps}{\texttt{CAPS}\xspace}
\newcommand{\cvfreco}{\texttt{CVFR-ECO}\xspace}
\newcommand{\cvfrhq}{\texttt{CVFR-HQ}\xspace}
\newcommand{\cvfr}{\texttt{CVFR}\xspace}

\newcommand{\ES}{$E_{\text{S}}$}
\newcommand{\LS}{$L_{\text{S}}$}
\newcommand{\hS}{$h_{\text{S}}$}

\newcommand{\BDRP}{BDR\textsubscript{P}}
\newcommand{\BDRV}{BDR\textsubscript{V}}

\newcommand{\vJ}{$v_{\text{J}}$}

\usepackage{lipsum} 
\usepackage{tikz,xcolor,hyperref}
\definecolor{lime}{HTML}{A6CE39}
\DeclareRobustCommand{\orcidicon}{%
	\begin{tikzpicture}
	\draw[lime, fill=lime] (0,0) 
	circle [radius=0.16] 
	node[white] {{\fontfamily{qag}\selectfont \tiny ID}};
	\draw[white, fill=white] (-0.0625,0.095) 
	circle [radius=0.007];
	\end{tikzpicture}
	\hspace{-2mm}
}

\foreach \x in {A, ..., Z}{%
	\expandafter\xdef\csname orcid\x\endcsname{\noexpand\href{https://orcid.org/\csname orcidauthor\x\endcsname}{\noexpand\orcidicon}}
}

\begin{document}

\title{Content-Adaptive Variable Framerate Encoding Scheme for Green Live Streaming}

\author{Vignesh V Menon\orcidA{}, \IEEEmembership{(Student Member,~IEEE)}, Samira Afzal\orcidB{}, \IEEEmembership{(Member,~IEEE)},\\ Prajit T Rajendran\orcidC{}, \IEEEmembership{(Student Member,~IEEE)}, Klaus Schoeffmann\orcidD{}, \IEEEmembership{(Member,~IEEE)},\\ Radu Prodan\orcidF{}, and Christian Timmerer\orcidE{}, \IEEEmembership{(Senior Member,~IEEE)}

\thanks{We acknowledge the financial support of the Austrian Federal Ministry for Digital and Economic Affairs, the National Foundation for Research, Technology and Development, and the Christian Doppler Research Association. Christian Doppler Laboratory ATHENA: \url{https://athena.itec.aau.at/}.}
\thanks{Vignesh V Menon is with the Video Coding Systems research group at the Video Communication and Applications department, Fraunhofer HHI, Berlin (e-mail: vignesh.menon@hhi.fraunhofer.de).}
\thanks{Samira Afzal, Klaus Schoeffmann, Radu Prodan and Christian Timmerer are with the Institute of Information Technology, Alpen-Adria-Universität Klagenfurt, Austria (e-mail: samira.afzal@aau.at, klaus.schoeffmann@aau.at, radu.prodan@aau.at, christian.timmerer@aau.at).}
\thanks{Prajit T Rajendran is with CEA, List, F-91120 Palaiseau, Université Paris-Saclay, France (e-mail: prajit.thazhurazhikath@cea.fr).}
}


\maketitle

\begin{abstract}
Adaptive live video streaming applications use a fixed predefined configuration for the bitrate ladder with constant framerate and encoding presets in a session. However, selecting optimized framerates and presets for every bitrate ladder representation can enhance perceptual quality, improve computational resource allocation, and thus, the streaming energy efficiency. In particular, low framerates for low-bitrate representations reduce compression artifacts and decrease encoding energy consumption. In addition, an optimized preset may lead to improved compression efficiency. To this light, this paper proposes a \underline{C}ontent-adaptive \underline{V}ariable \underline{F}rame\underline{r}ate (\cvfr) encoding scheme, which offers two modes of operation: \textit{ecological} (\texttt{ECO}) and \textit{high-quality} (\texttt{HQ}). \cvfreco optimizes for the highest encoding energy savings by predicting the optimized framerate for each representation in the bitrate ladder. \cvfrhq takes it further by predicting each representation's optimized framerate-encoding preset pair using low-complexity \DCT{} energy-based spatial and temporal features for compression efficiency and sustainable storage. We demonstrate the advantage of \cvfr using the x264 open-source video encoder. The results show that \cvfreco yields an average PSNR and VMAF increase of \SI{0.02}{\decibel} and 2.50 points, respectively, for the same bitrate, compared to the fastest preset highest framerate encoding. \cvfreco also yields an average encoding and storage energy consumption reduction of \SI{34.54}{\percent} and \SI{76.24}{\percent} considering a \jnd (JND) of six VMAF points. In comparison, \cvfrhq yields an average increase in PSNR and VMAF of \SI{2.43}{\decibel} and 10.14 points, respectively, for the same bitrate. Finally, \cvfrhq resulted in an average reduction in storage energy consumption of \SI{83.18}{\percent} considering a JND of six VMAF points.
\end{abstract}


\input{sec_intro}
\input{sec_vfr}

\input{sec_zero_latency}
\input{sec_proposed_method}

\input{sec_experimental_design}
\input{sec_evaluation}
\input{sec_conclusions}

\bibliography{references.bib}{}
\bibliographystyle{IEEEtran}

\begin{IEEEbiography}[{\includegraphics[width=1.0in,height=1.25in,clip,keepaspectratio]{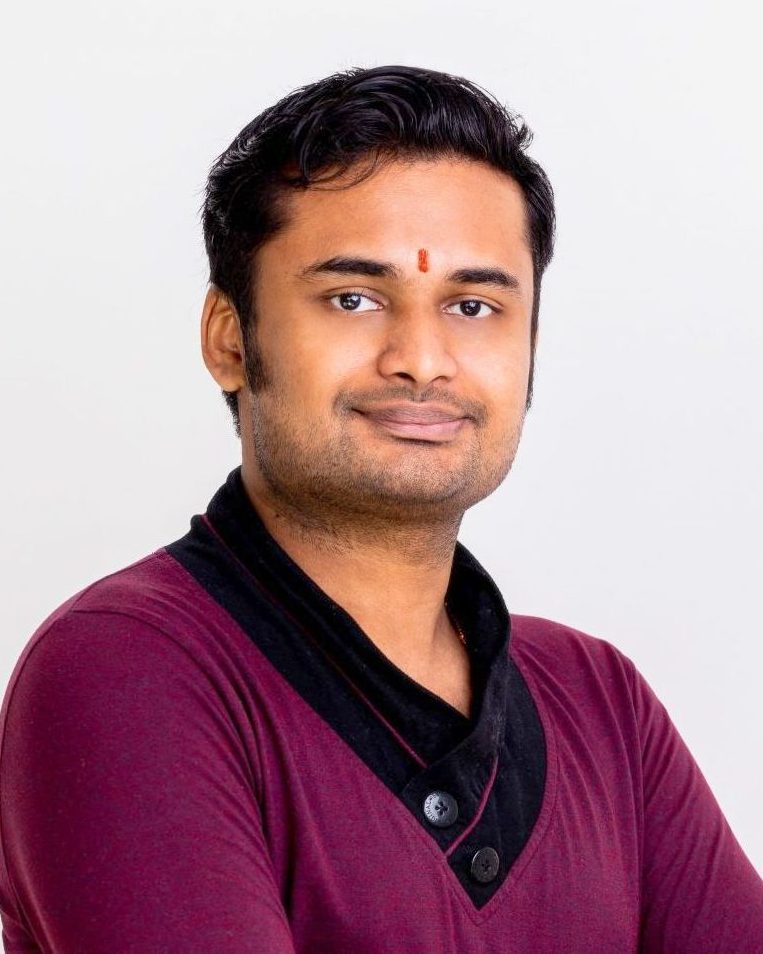}}]
{Vignesh V Menon} is a researcher at Fraunhofer HHI, Berlin, and a Ph.D. candidate at the Institute of Information  Technology (ITEC), Alpen-Adria-Universität Klagenfurt (AAU). He received a B.Tech. in Electronics and Communication Engineering from Amrita Vishwa Vidyapeetham University, India, and an M.Sc. in Information and Network Engineering from KTH Royal Institute of Technology, Sweden, in 2016 and 2020, respectively. He was a software engineer developing video encoding software solutions in MulticoreWare Inc., India, between 2016-2018 and Divideon, Sweden, between 2018-2020. His research interests are video streaming, image, and video compression. Further information at \url{https://vigneshvijay94.com}.
\end{IEEEbiography}
\vspace{-20 pt}
\begin{IEEEbiography}[{\includegraphics[width=1.0in,height=1.25in,clip,keepaspectratio]{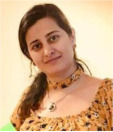}}]
{Samira Afzal} is a Postdoctoral researcher at Alpen-Adria-Universität Klagenfurt (AAU), currently working on the GAIA project. She received her Ph.D. in Electrical Engineering at the School of Electrical and Computer Engineering (FEEC), University of Campinas (UNICAMP), in November 2019. She received her master’s degree in IT engineering at Sharif University of Technology (SUT) in 2011.
\end{IEEEbiography}
\vspace{-20 pt}
\begin{IEEEbiography}[{\includegraphics[width=1.0in,height=1.25in,clip,keepaspectratio]{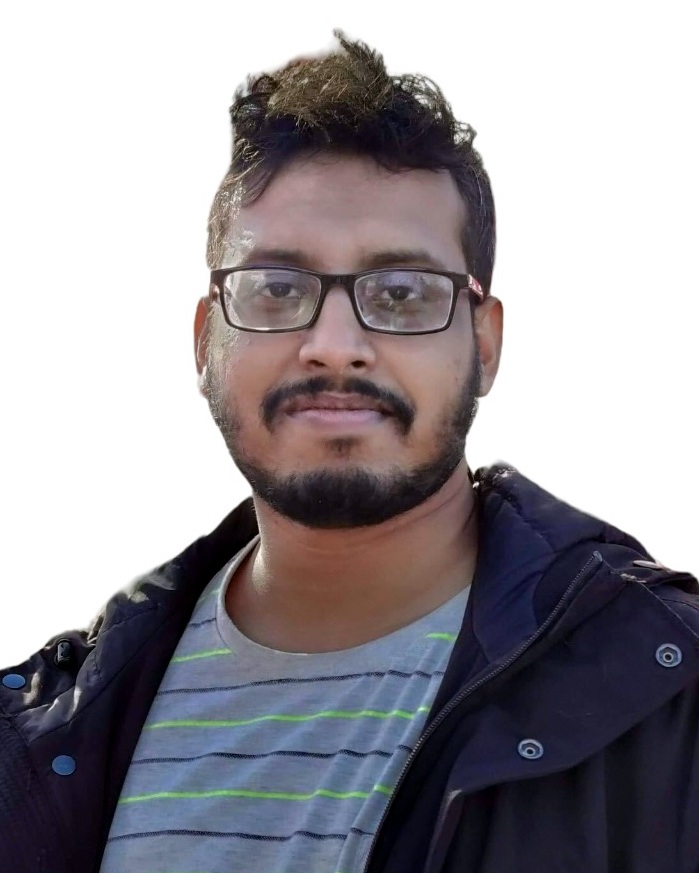}}]
{Prajit T Rajendran} is a Ph.D. candidate at Université Paris-Saclay, working on his doctoral research project in collaboration with CEA LIST and LNE France. He received his B.Engg degree in Electronics and Communication Engineering from Ramaiah Institute of Technology, Bangalore, India, in 2018 and subsequently an M.Sc. degree in Information and Network Engineering from KTH Royal Institute of Technology, Sweden, in 2020. His research interests include computer vision, deep learning, active learning, and human-in-the-loop artificial intelligence.
\end{IEEEbiography}
\vspace{-20 pt}
\begin{IEEEbiography}[{\includegraphics[width=1.0in,height=1.25in,keepaspectratio]{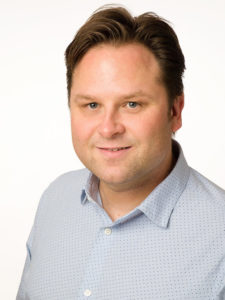}}]{Klaus Schoeffmann} is an associate professor at the Institute of Information Technology (ITEC). He received his Ph.D. degree and Habilitation (venia docendi) from Klagenfurt University in both 2009 and 2015, respectively, in computer science. He is currently an Associate Professor with the Institute of Information Technology (ITEC), Klagenfurt University, Klagenfurt, Austria. His research focuses on video analytics and interactive multimedia systems, particularly in medicine. He has co-authored more than 110 publications on various topics in multimedia. He has co-organized several international conferences, workshops, and special sessions in the field of multimedia. Furthermore, he is a co-founder of the Video Browser Showdown (VBS), a member of the ACM, and a regular reviewer for international conferences and journals in multimedia. \end{IEEEbiography}
\vspace{-20 pt}
\begin{IEEEbiography}[{\includegraphics[width=1.0in,height=1.25in,keepaspectratio]{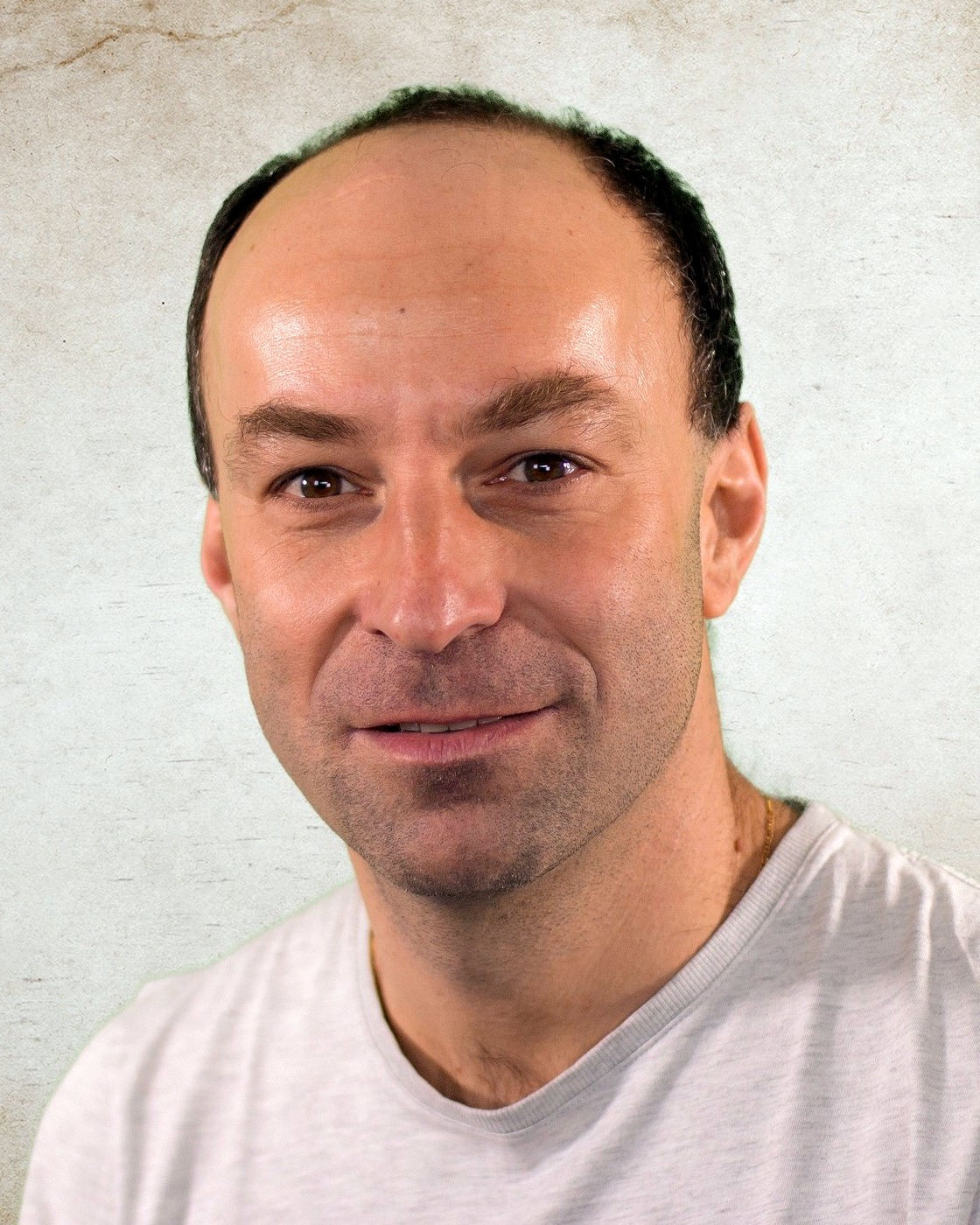}}]{Radu Prodan} is a professor in distributed systems at ITEC, University of Klagenfurt, Austria. Previously, he was an associate professor at the University of Innsbruck, Austria. He received his Ph.D. in 2004 from the Vienna University of Technology. His research interests are performance, optimization, and resource management tools for parallel and distributed systems. He participated in numerous projects and coordinated the European Union projects ARTICONF and Graph-Massivizer. He co-authored over 200
publications and received three IEEE best paper awards. \end{IEEEbiography}
\vspace{-20 pt}
\begin{IEEEbiography}[{\includegraphics[width=1.0in,height=1.25in,clip,keepaspectratio]{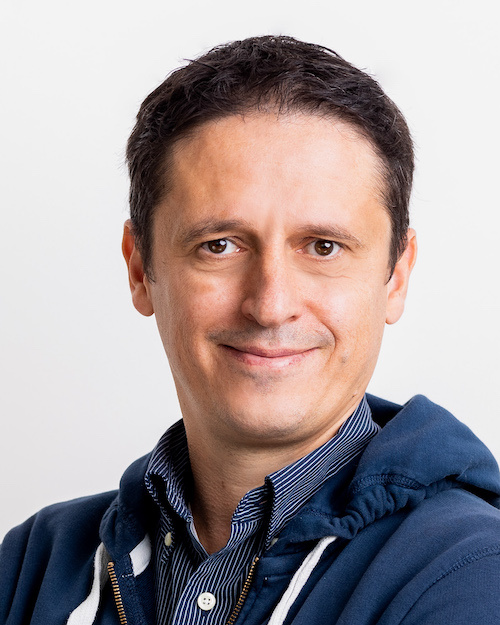}}]
{Christian Timmerer}(M'08-SM'16) is a full professor of computer science at Alpen-Adria-Universität Klagenfurt (AAU), Institute of Information Technology (ITEC), and he is the director of the Christian Doppler (CD) Laboratory ATHENA (\url{https://athena.itec.aau.at/}). His research interests include multimedia systems, immersive multimedia communication, streaming, adaptation, and quality of experience, where he co-authored seven patents and more than 300 articles. He was the general chair of WIAMIS 2008, QoMEX 2013, MMSys 2016, and PV 2018 and has participated in several EC-funded projects, notably DANAE, ENTHRONE, P2P-Next, ALICANTE, SocialSensor, COST IC1003 QUALINET, ICoSOLE, and SPIRIT. He also participated in ISO/MPEG work for several years, notably in MPEG-21, MPEG-M, MPEG-V, and MPEG-DASH, where he served as standard editor. In 2012, he cofounded Bitmovin (\url{http://www.bitmovin.com/}) to provide professional services around MPEG-DASH, where he holds the position of the Chief Innovation Officer (CIO) –- Head of Research and Standardization. Further information at \url{http://timmerer.com}. 
\end{IEEEbiography}
\balance
\end{document}

%% file: sec_intro.tex
\section{Introduction}

\IEEEPARstart{S}{treaming} video has become the most popular online activity, with viewers accessing content across various devices such as TVs, laptops, tablets, and smartphones~\cite{Cisco_ref}. Optimized delivery requires video encoding into multiple quality levels adjusted to network conditions and user devices~\cite{DASH_ref}. \HAS (HAS) has become the \textit{de-facto} standard in delivering video content for various clients regarding internet speeds and device types. HAS divides the video content into segments and encodes each segment at various bitrates and resolutions, called \textit{representations}, stored in plain HTTP servers, which continuously adapt the video delivery to the network conditions and device capabilities of the client~\cite{DASH_ref}. In the current streaming landscape, \emph{live-streaming} is one of the foremost challenges, which necessitates faster compression and simplified encoding techniques~\cite{bitmovin_devel_report_ref} to reduce the time between video capture and playback~\cite{livestream_ref}. The viewer's experience improves by the reduced latency, enabling a more responsive and engaging content consumption~\cite{zero_latency_ref, zero_latency_ref2}. 

\subsubsection{Low-latency live-streaming}
A low-latency live encoder must maintain an encoding speed greater than the video framerate regardless of the complexity of the video content. A reduced encoding speed can lead to frames dropping during transmission~\cite{pradeep_ref}, which may decrease the QoE. Traditionally, live-streaming sessions use a fixed bitrate ladder, such as \HLS (HLS)~\cite{HLS_ladder_ref}. Per-title encoding schemes that optimize bitrate ladder according to the specific video content~\cite{netflix_paper, pte_ref2} received wide adoption for only \VOD (VoD) services due to the expensive \textit{convex-hull} computation~\cite{gnostic, jtps_ref}. Moreover, state-of-the-art per-title encoding schemes optimize target resolution~\cite{netflix_paper, pte_ref2, gnostic, opte_ref} or bitrate~\cite{jtps_ref} based on perceptual quality. Noteworthy, these approaches are not suitable for live-streaming solutions, where (i) predictable switching between representations and (ii) simplified player logic on the client side are of significant importance~\cite{livestream_ref}. Hence, per-title encoding optimization using fixed bitrate-resolution pairs is essential.

\subsubsection{Energy consumption}
Bitrate ladder encoding for HAS platforms incurs substantial energy consumption, thereby straining environmentally conscious resource management in data centers~\cite{koziri2018efficient}. Encoding energy optimization in this context aims to reduce the ecological footprint and operational costs associated with adaptive live-streaming~\cite{carbontrust2021}. Additionally, storing the video on the server and streaming it to devices using content distribution networks (CDNs) consume energy~\cite{cdn_farahani, bianco2016energy, carbontrust2021}. Encoding solutions in video streaming directly impact storage and transmission energy consumption. Consequently, minimizing the overall energy consumption in video streaming is a significant challenge in the industry today~\cite{carbontrust2021, nokia_ref,adobe_ref}. Thus, energy-efficient encoding techniques are necessary to reduce overall consumption without compromising the perceptual quality of the delivered video content.

\subsubsection{Low-latency, minimized perceptual redundancy}
\begin{figure}[t]
\centering
    \includegraphics[width=0.492\textwidth]{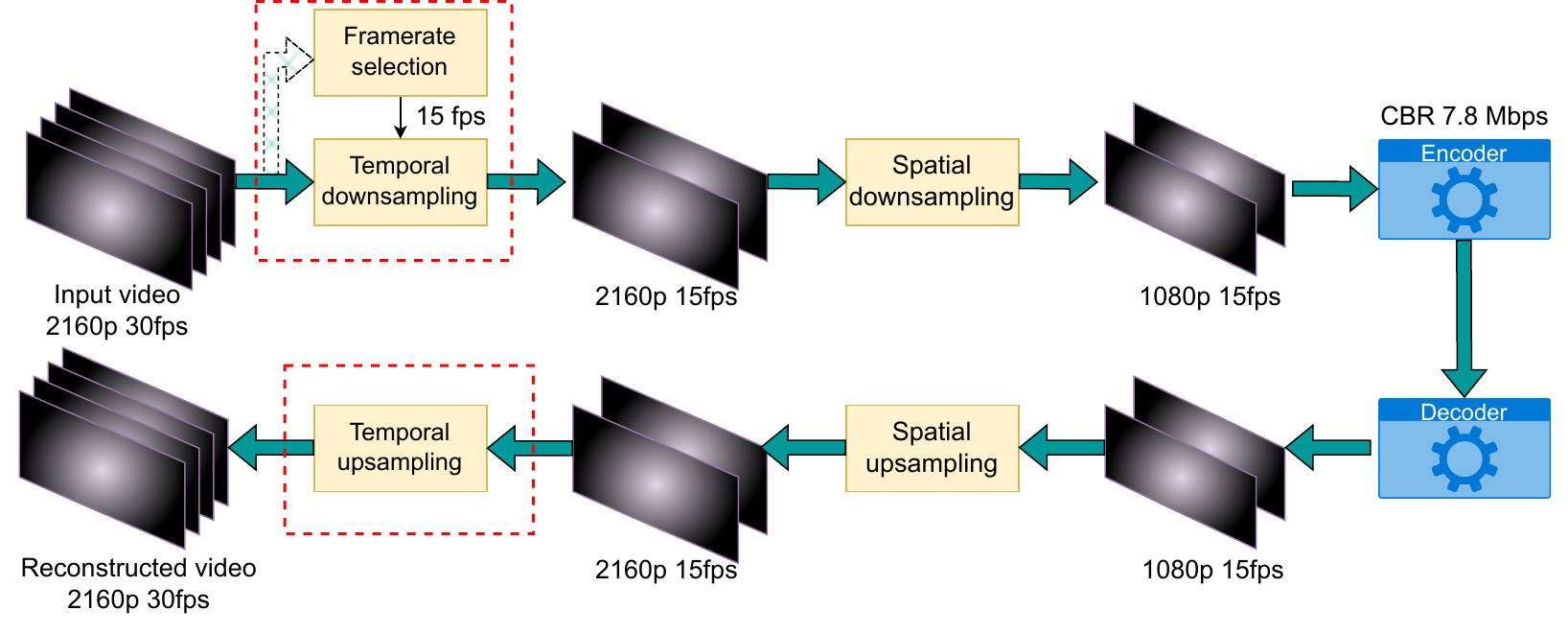}
\caption{An example scenario of a VFR encoding scheme for adaptive streaming that encodes a video segment of ultra-high definition resolution and 30fps in representation (7.8 Mbps, 1080p) with the selected framerate of 15fps. On the client side, red dashed blocks indicate the additional steps introduced compared to the traditional bitrate ladder encoding.}
\label{fig:vfr_arch}
\end{figure}

This paper targets a low-latency encoding scheme yielding optimized trade-offs between overall energy consumption and compression efficiency without significantly changing the streaming architecture. To achieve this goal, we consider \emph{variable framerate} (VFR) encoding methods~\cite{vfrc_ref, vfr_csvt1_ref, vfr_csvt2_ref, vfr_pcs1_ref} introduced in the literature to limit the modifications to the encoding server and client. 
This approach enhances adaptive streaming mechanisms by aligning the framerate selection with perceptual optimization, allowing for dynamic framerate adjustment based on network conditions and viewer devices. Fig.~\ref{fig:vfr_arch} shows an example of VFR encoding for streaming applications that temporarily downsamples the video of the original framerate of 30\,fps to a framerate of 15\,fps indicated by the framerate selection module. After decoding, the video is temporally upscaled to its original framerate. Noteworthy, state-of-the-art video encoders offer several \textit{presets} to balance the trade-off between encoding time and compression efficiency~\cite{caps_ref} that differ in the encoding tools used. Faster presets utilize a subset of the tools to reduce encoding time and energy consumption~\cite{Silveira2017Performance}. Live encoders usually choose faster presets to encode video frames in real-time, sacrificing some quality and compression ratio. Hence, this paper also uses variable preset encoding methods in conjunction with VFR to optimize perceptual quality and energy consumption. 
Furthermore, a \jnd (JND)-aware representation elimination optimizes the allocation of streaming bits based on the perceptual thresholds of human vision~\cite{jnd_ref}. Ensuring that the adjacent points of the bitrate ladder have a perceptual quality difference of at least one JND eliminates the perceptual redundancy between representations and reduces the overall energy consumption. This paper considers JND as a function of VMAF~\cite{VMAF}, and future work will study other functions. 

\subsubsection{Contributions}
This paper contains five contributions.
\paragraph{Comprehensive compression efficiency and energy consumption analysis} uses a state-of-the-art video encoder at multiple framerates and encoding presets. This highlights the complexity of this relationship and the need for adaptive solutions that balance video quality and energy efficiency.
\paragraph{\cvfr} \emph{content-adaptive variable framerate encoding} scheme optimizes the bitrate ladder by maximizing the perceptual quality and energy efficiency while maintaining the target encoding speed for low-latency encoding. A JND-based representation elimination algorithm removes the perceptual redundancy in the bitrate ladder. \cvfr offers two operational modes: \emph{ecological} (\cvfreco) and \emph{high-quality} (\cvfrhq).
\paragraph{\cvfreco} predicts the optimized framerate for the fastest encoding preset for each representation of the bitrate ladder, yielding the lowest encoding energy consumption. 
\paragraph{\cvfrhq} jointly predicts the optimized framerate-encoder preset pairs for each representation to yield the highest possible compression efficiency while maintaining the target encoding speed for low-latency encoding. 
\paragraph{Experimental evaluation} compares the \cvfr schemes with state-of-the-art encoding methods in terms of compression efficiency and energy consumption.  \cvfreco reduces encoding and storage energy consumption by \SI{34.54}{\percent} and \SI{76.24}{\percent} during adaptive live video streaming, respectively. On the other hand, \cvfrhq significantly improves compression efficiency, achieving a PSNR increase of \num{2.43} dB and VMAF increase of \num{10.14} points for the same bitrate. Additionally, \cvfrhq reduces storage energy consumption by \SI{83.18}{\percent} during adaptive live-streaming.


%% file: sec_vfr.tex
\section{Variable framerate encoding}
\label{sec:vfr}
Several studies highlighted the effectiveness of raising the framerate in reducing temporal artifacts, including flickering, stuttering, and motion blur~\cite{qoe6,LFR,MG}. Encoding each frame involves various computational operations, including motion estimation, transformation, quantization, and entropy coding. Higher framerates lead to more frames processed per unit of time, increased computational workload, and higher energy consumption during encoding~\cite{alaoui2014energy}. Moreover, encoding at higher framerates might require higher bitrates to maintain video quality, potentially increasing energy use. Sometimes, the trade-off between encoding speed, energy consumption, and video quality might suggest a more energy-efficient encoding at a lower framerate for a given quality ~\cite{alaoui2014energy,vfrc_ref}.  

\begin{figure*}[t]
\centering
\begin{subfigure}{0.235\textwidth}
    \centering
    \includegraphics[clip,width=\textwidth]{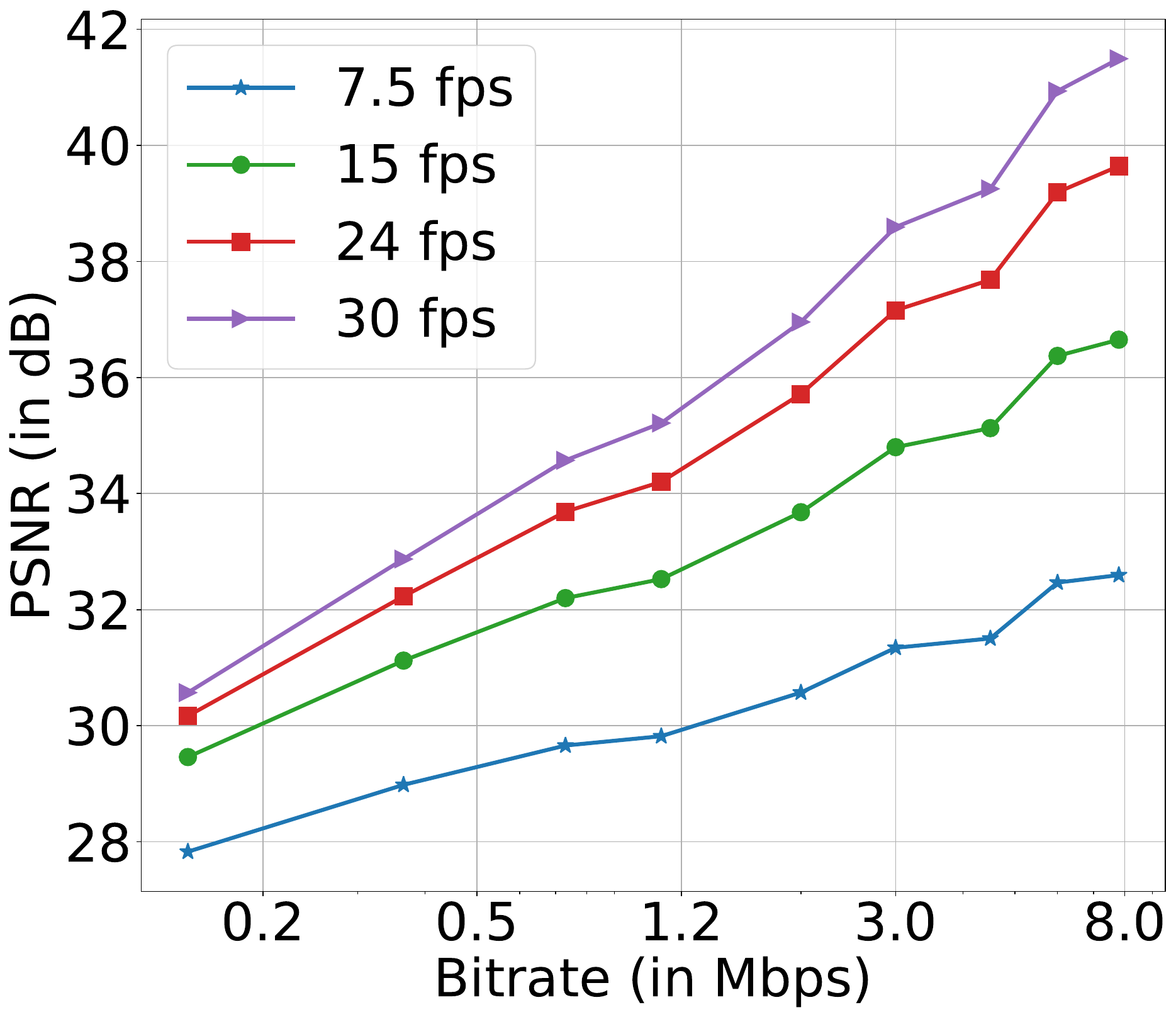}
    \caption{Average PSNR.}
    \label{fig:fast_vfr_bitrate_psnr_plot}
\end{subfigure}
\hfill
\begin{subfigure}{0.235\textwidth}
    \centering
    \includegraphics[clip,width=\textwidth]{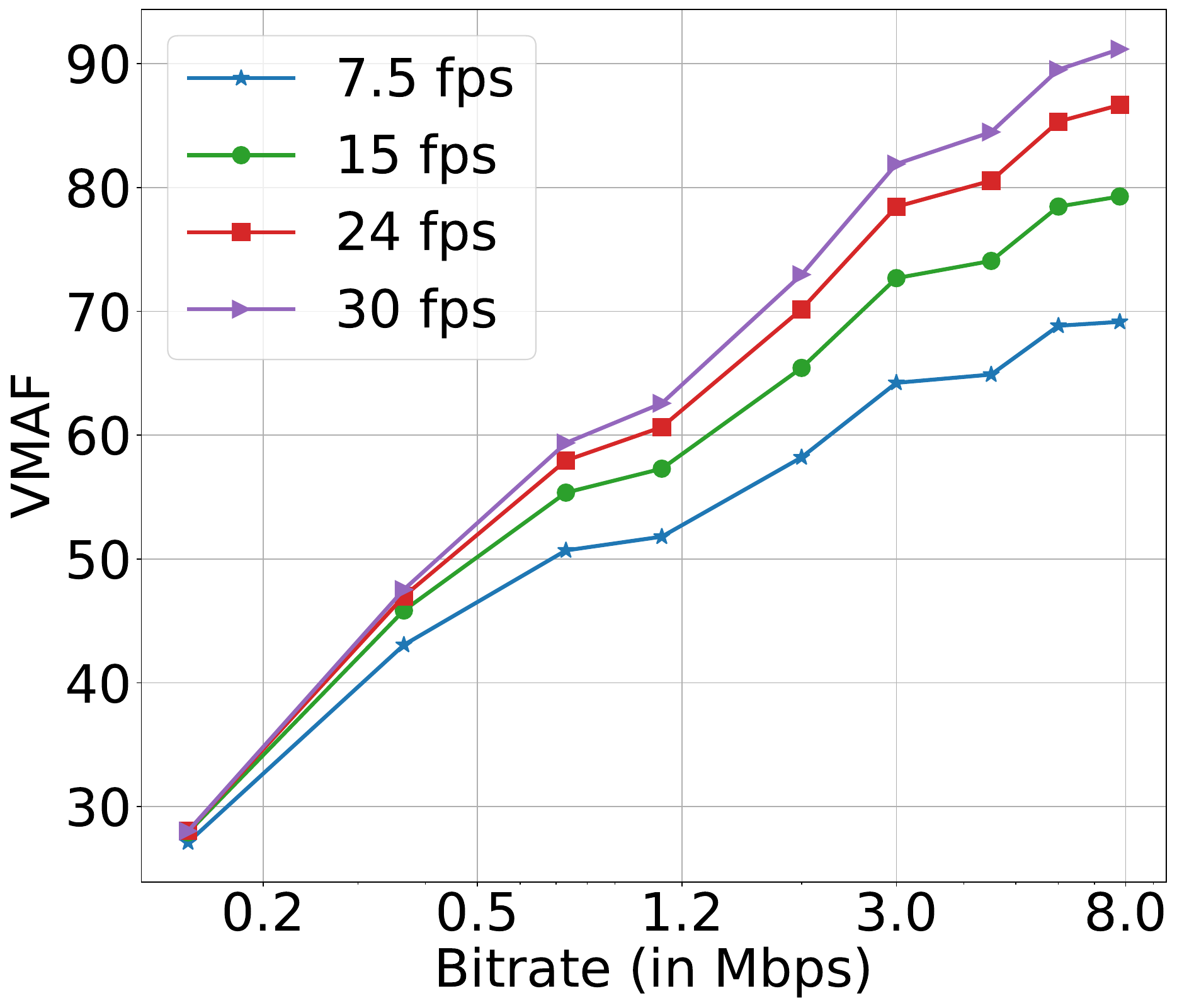}
    \caption{Average VMAF.}
    \label{fig:fast_vfr_bitrate_VMAF_plot}
\end{subfigure}
\begin{subfigure}{0.235\textwidth}
    \centering
    \includegraphics[clip,width=\textwidth]{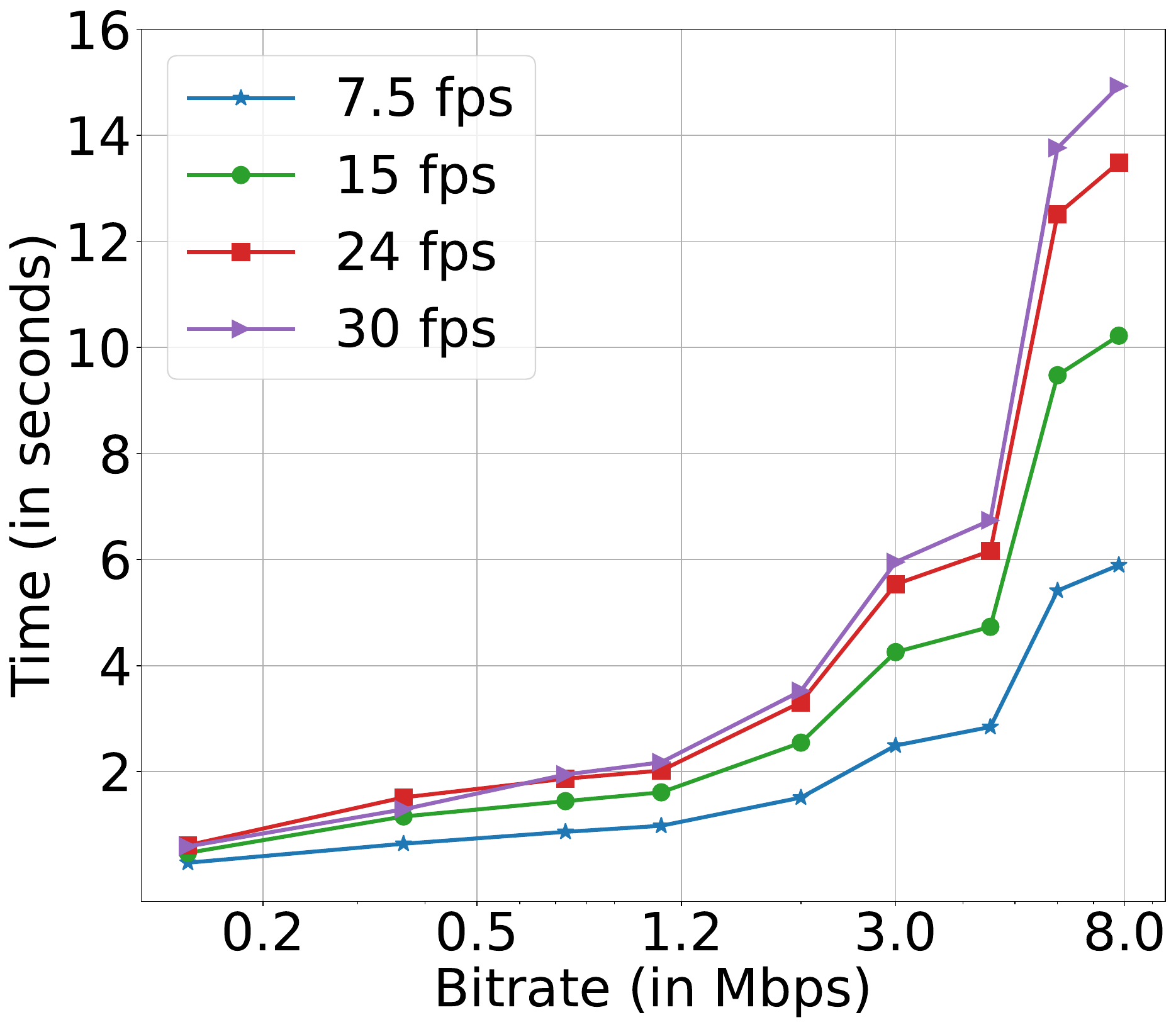}
    \caption{Average encoding time.}
    \label{fig:fast_vfr_bitrate_time_plot}
\end{subfigure}
\hfill
\begin{subfigure}{0.245\textwidth}
    \centering
    \includegraphics[clip,width=\textwidth]{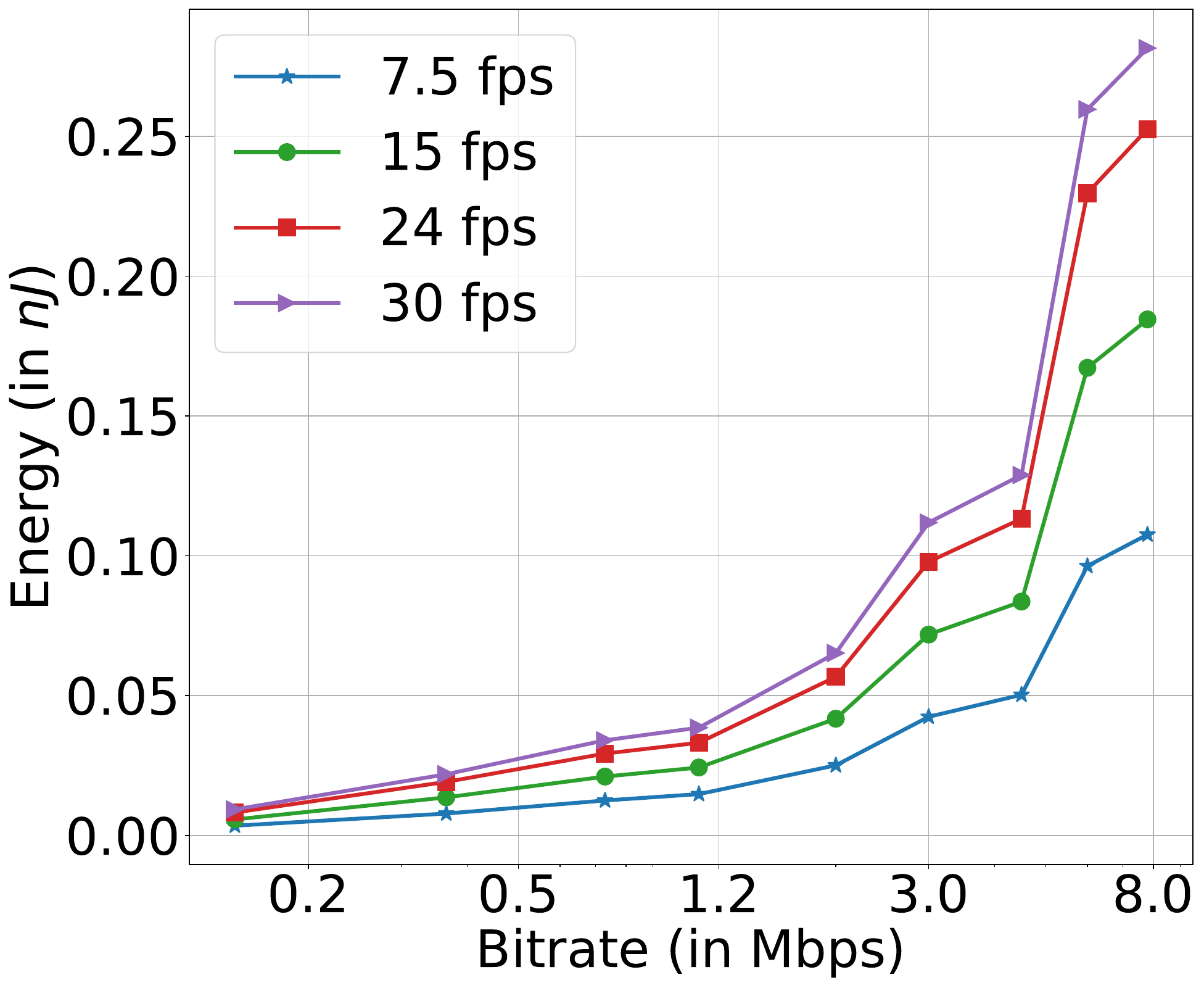}
    \caption{Average energy consumption.}
    \label{fig:fast_vfr_bitrate_energy_plot}
\end{subfigure}
\caption{Average encoding metrics for 7.5 fps, 15 fps, 24 fps, and 30 fps HLS CBR encoding using the \textit{veryslow} preset of the x264 AVC encoder~\cite{x264_ref, AVC} using VCD dataset~\cite{VCD_ref}.}
\label{fig:fast_vfr_intro_plots}
\end{figure*}

\begin{table*}[t]
\caption{Comparison of the state-of-the-art VFR encoding methods with \cvfr.}
\centering
\begin{tabular}{l||c|c|c|c}
\specialrule{.12em}{.05em}{.05em}
\specialrule{.12em}{.05em}{.05em}
Method & Target scenario & Framerate estimation method & Number of pre-encodings & Scalability\\
\specialrule{.12em}{.05em}{.05em}
\specialrule{.12em}{.05em}{.05em}
Bruteforce~\cite{netflix_paper} & VoD & Bruteforce encoding & $\Tilde{b}\times \Tilde{f}$ & Yes \\
Huang~\etal~\cite{vfr_tb1_ref} & VoD & Prediction using support vector regression method & 0 & Yes \\
Katsenou~\etal~\cite{vfr_pcs1_ref} & VoD & Prediction using bagged decision trees & 0 & No \\
ViSTRA~\cite{vfr_csvt2_ref} & VoD & Encoding at two framerates & 2 & No \\
Herrou~\etal~\cite{vfrc_ref} & Live & Prediction using random forest classifier models & 0 & No \\
\specialrule{.12em}{.05em}{.05em}
\cvfr & Live & Prediction using random forest regression models & 0 & Yes\\
\specialrule{.12em}{.05em}{.05em}
\specialrule{.12em}{.05em}{.05em}
\end{tabular}
\label{tab:vfr_sota}
\end{table*}

\subsection{Motivation} Mackin~\etal~\cite{vfrc_ref1} found that higher framerates are more encoding efficient at higher bitrates, particularly in simple sequences with camera movements.
This content dependency to select the optimized framerate is the basis for a VFR coding scheme. Accordingly, the example scenario in Fig.~\ref{fig:vfr_arch} and the results in Fig.~\ref{fig:fast_vfr_bitrate_psnr_plot} and~\ref{fig:fast_vfr_bitrate_VMAF_plot}, demonstrate a similar trend in the average  PSNR and VMAF metrics for 7.5 fps, 15 fps, 24 fps, and 30 fps encodings of the VCD dataset~\cite{VCD_ref} using the \texttt{veryslow} preset of the x264 encoder~\cite{x264_ref}. The assumption is that dropping frames in slow-motion videos has a less noticeable difference in the perceived moving objects' quality than in fast-motion. This coding scheme can considerably reduce the bitrate and encoding energy without incurring apparent distortions.
Additionally, a higher framerate increases the encoding time (\cf Fig.~\ref{fig:fast_vfr_bitrate_time_plot}) and energy (\cf Fig.~\ref{fig:fast_vfr_bitrate_energy_plot}) overheads measured using the \texttt{CodeCarbon} tool for the x264 encoder on an Intel Xeon Gold 5218R processor using eight threads.  Finally, the average perceptual quality difference (measured using VMAF~\cite{VMAF}) between multiple framerate encodings at low bitrates is insignificant. Closer quality for different framerates at low bitrates could be due to the perceptual sensitivity of the VMAF metric. Encoding decisions at low bitrates may prioritize certain video aspects, leading to similar perceptual quality even for different framerates.

\subsection{State-of-the-art VFR scheme architecture}
State-of-the-art VFR schemes include a temporal downsampling step before encoding and a temporal upsampling step after decoding (\cf Fig.~\ref{fig:vfr_arch}), described in this section. 

\subsubsection{Temporal downsampling} involves discarding or reducing the number of frames in the video to achieve a lower framerate, reducing the temporal detail and smoothness of the video. There are two standard downsampling techniques. 
\paragraph{Frame dropping} intentionally discards or drops specific frames from the video sequence based on a regular pattern (\eg every $n$\textsuperscript{th} frame) or dynamically determined based on specific criteria (\eg low visual importance)~\cite{drop_ref1}.
\paragraph{Temporal filtering} analyzes the temporal redundancy between frames and applies filtering or motion analysis to generate new frames that are blends or interpolations of adjacent frames. The resulting frames can reduce the framerate while maintaining smoother motion and minimizing judder or jerkiness in the video.

\begin{figure*}[t]
\centering
\begin{subfigure}{0.235\textwidth}
    \centering
    \includegraphics[clip,width=\textwidth]{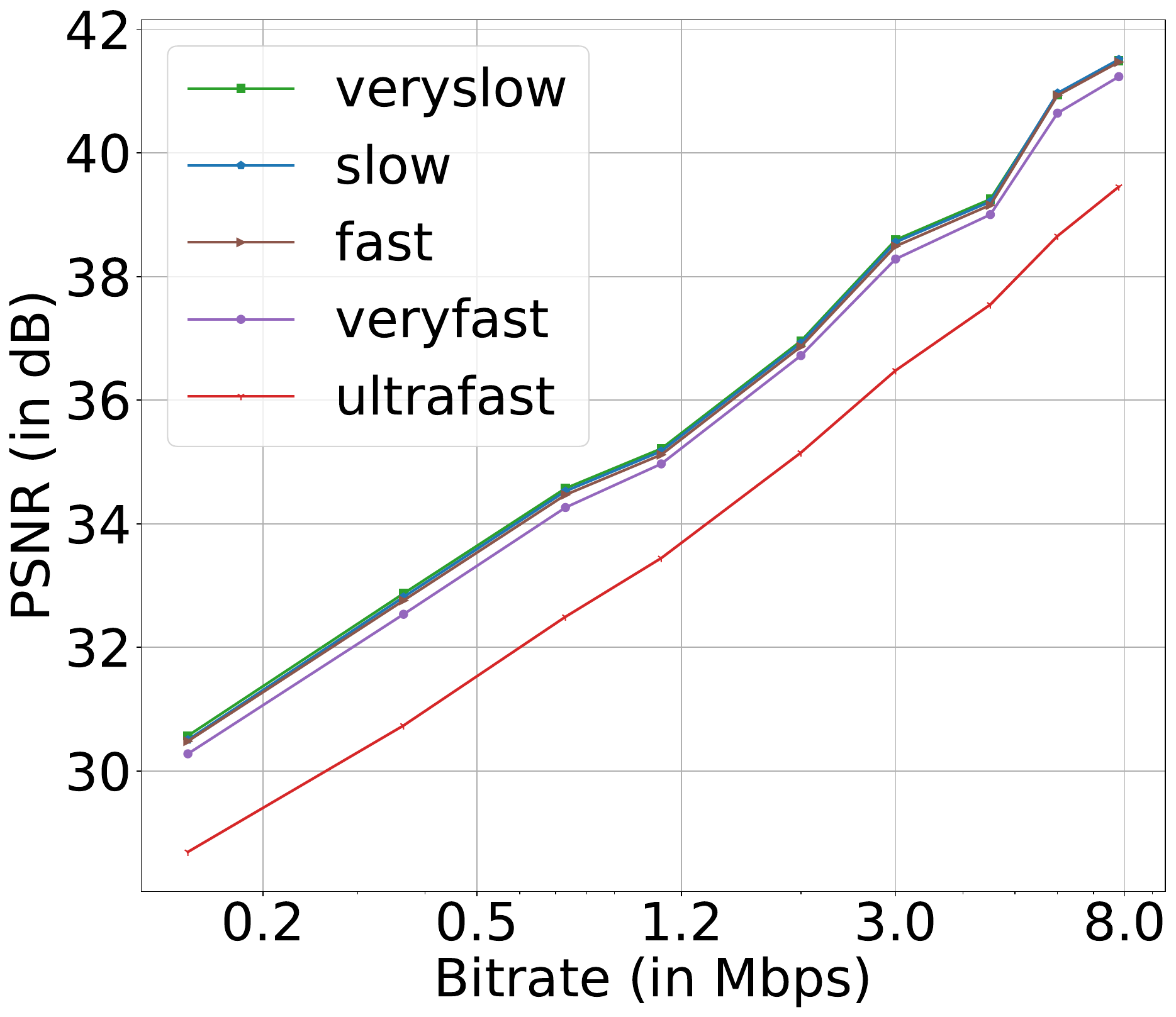}
    \caption{Average PSNR.}
    \label{fig:30fps_bitrate_psnr_plot}
\end{subfigure}
\hfill
\begin{subfigure}{0.235\textwidth}
    \centering
    \includegraphics[clip,width=\textwidth]{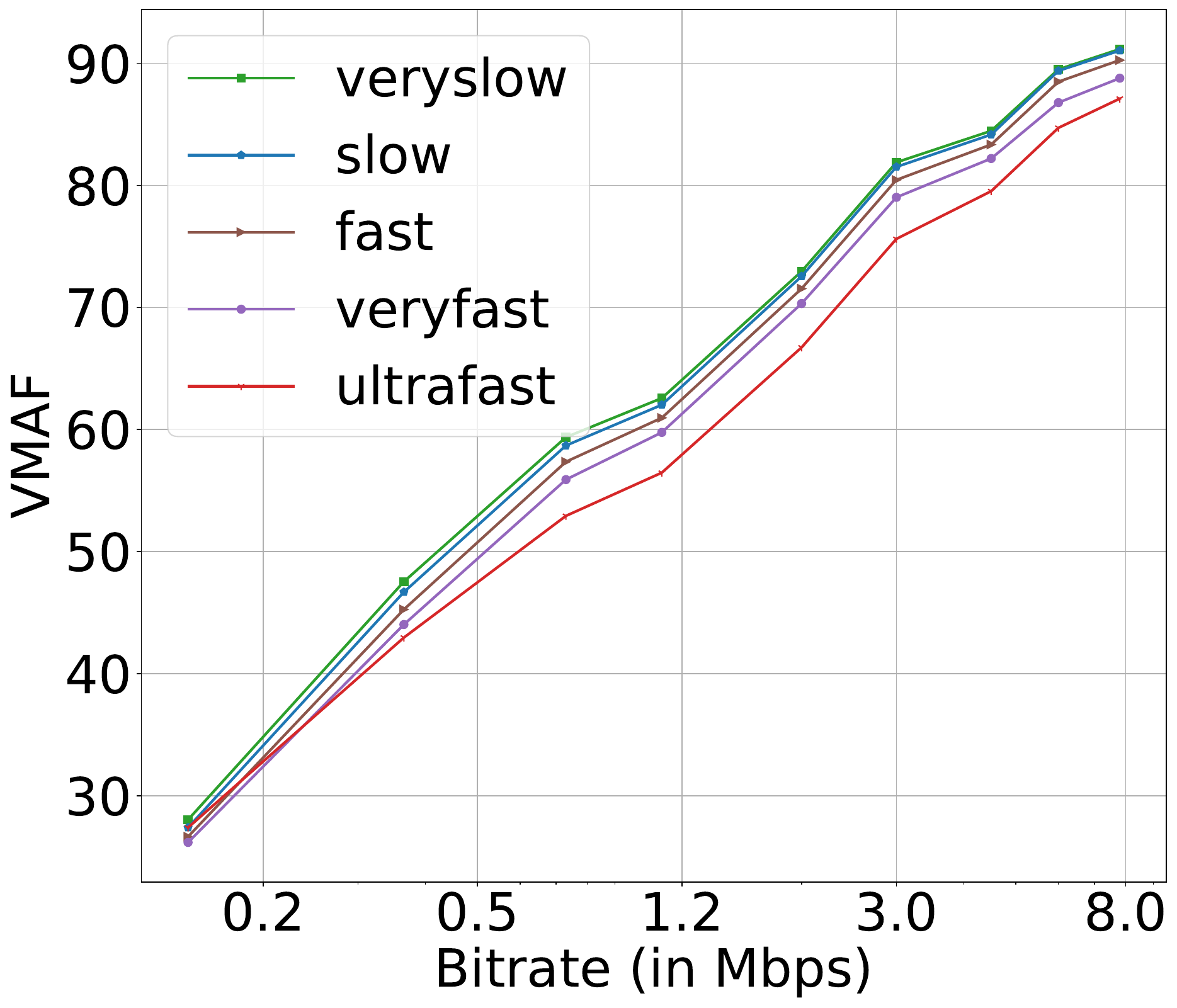}
    \caption{Average VMAF.}
    \label{fig:30fps_bitrate_VMAF_plot}
\end{subfigure}
\begin{subfigure}{0.235\textwidth}
    \centering
    \includegraphics[clip,width=\textwidth]{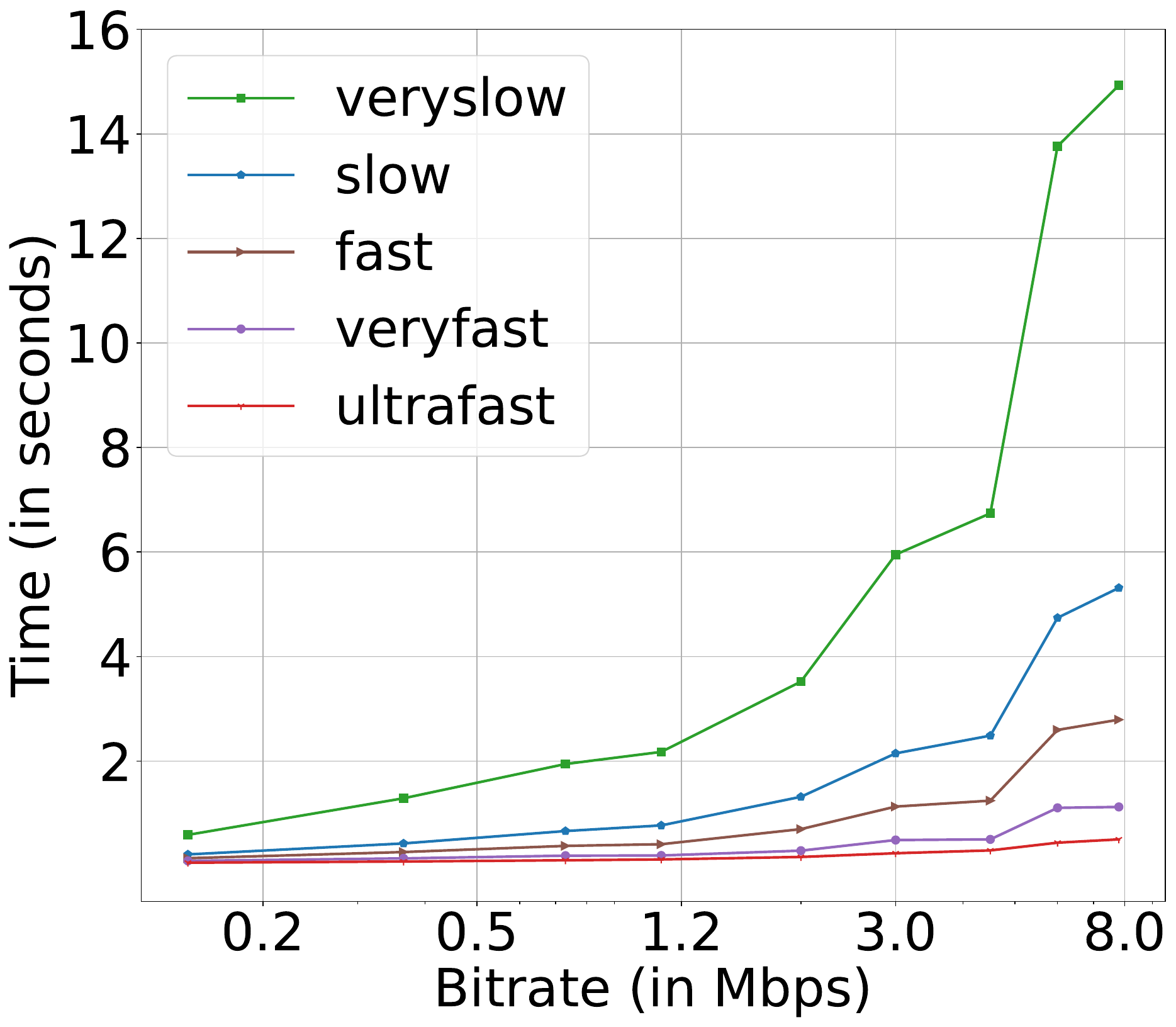}
    \caption{Average encoding time.}
    \label{fig:30fps_bitrate_time_plot}
\end{subfigure}
\hfill
\begin{subfigure}{0.245\textwidth}
    \centering
    \includegraphics[clip,width=\textwidth]{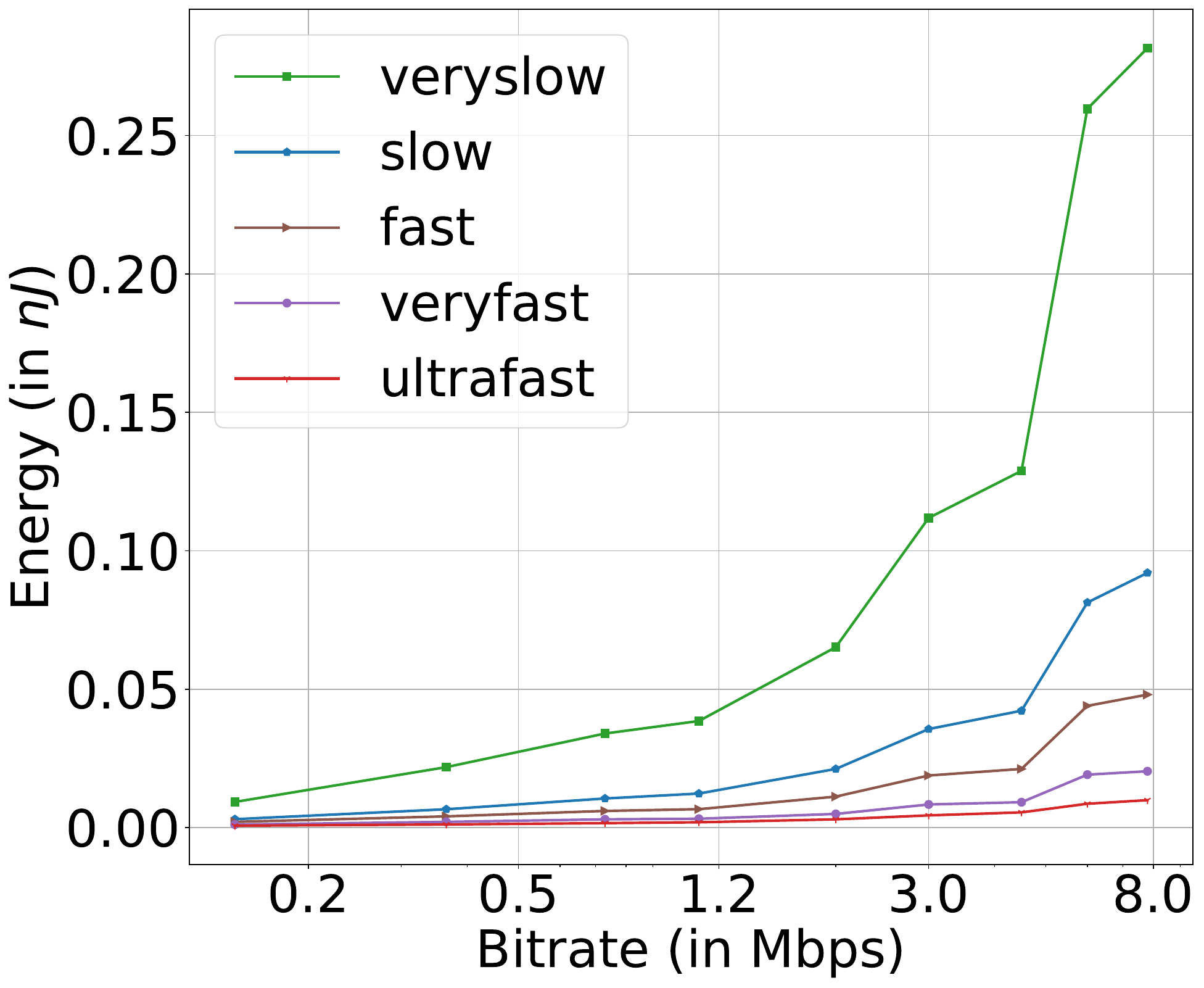}
    \caption{Average energy consumption.}
    \label{fig:30fps_bitrate_energy_plot}
\end{subfigure}
\caption{Average encoding metrics for HLS CBR encoding at 30 fps using selected presets of x264 AVC encoder~\cite{x264_ref, AVC} using the VCD dataset~\cite{VCD_ref}.}
\label{fig:30fps_intro_preset_plots}
\end{figure*}

\subsubsection{Temporal upsampling} upsamples the VFR-encoded video to the original framerate at the client side, as most display devices do not support a variable framerate viewing~\cite{vfrc_ref}. There are two standard upsampling techniques.

\paragraph{Frame duplication} generates additional replicated frames of the already decoded ones and inserts them between consecutive frames. This straightforward technique does not require motion estimation or complex algorithms~\cite{coda_ref}. However, it provides no additional temporal information or motion details compared to the original video sequence.
\paragraph{Frame interpolation} generates new frames by estimating the intermediate motion between adjacent frames. Various algorithms, such as optical flow estimation~\cite{interpolation_ref3}, can estimate motion vectors and generate new frames~\cite{interpolation_ref2}. Frame interpolation can effectively increase the framerate, providing smoother motion in the video sequence~\cite{interpolation_ref1}.

\subsection{Related VFR works}
Many research works have investigated VFR. Some use different motion-related features with thresholding techniques~\cite{vfr_csvt2_ref} or machine learning algorithms~\cite{vfr_tb1_ref, vfr_pcs1_ref} to select the desired framerate. Table~\ref{tab:vfr_sota} shows the target scenario, framerate estimation method, and the number of pre-encodings needed to optimize the framerate of the state-of-the-art VFR methods. 
To determine the optimal framerate for $\Tilde{b}$ representations and $\Tilde{f}$ supported framerates, it is necessary to \textit{``bruteforce''} execute $\Tilde{b}\times \Tilde{f}$ test encodings~\cite{netflix_paper}. 

Huang~\etal \cite{vfr_tb1_ref} proposed a framerate selection mechanism to meet the ``satisfied user ratio'' using a support vector regression method. However, it uses complex and computationally expensive visual saliency and spatial randomness map features for each frame unsuitable for real-time dynamic framerates. 

Katsenou~\etal \cite{vfr_pcs1_ref} trained decision trees to predict the critical framerate at a sequence level using optical flow as the temporal and gray-level co-occurrence matrix as the spatial features. However, the feature extraction needs significant processing time, rendering it unfit for live-streaming applications.

ViSTRA~\cite{vfr_csvt2_ref} employs a temporal resolution optimization using a framerate-dependent quality metric~\cite{zhang_ref} to assess the perceptual quality difference between a temporally downsampled video frame and its full framerate original. This method involves encoding at the original framerate and half of the original framerate; hence, it introduces significant latency. 

Herrou~\etal \cite{vfrc_ref} proposed a VFR method to determine the minimum framerate that preserves the perceived video quality using two random forest classifiers. However, this method has limitations to three framerates (\ie original and downsampled by a factor of two and four). 

\subsection{Summary}
Most related works on VFR yield latency unsuitable for live-streaming applications. The methods use complex and computationally expensive features or pre-encodings, which introduce significant latency in streaming. The solution proposed in~\cite{vfrc_ref} is not \textit{scalable} and requires model retraining for different framerate and bitrate ladder representations. To mitigate these problems, this paper proposes a low-latency scalable solution using random-forest-based framerate prediction. Finally, due to the low computational complexity, a prerequisite for low-latency encoding, this paper considers frame dropping and frame duplication techniques for temporal downsampling and upsampling, respectively.

%% file: sec_zero_latency.tex
\section{Variable preset encoding}
\label{sec:zero_latency}
\subsection{Motivation}
Traditionally, video encoders provide predefined settings and configurations, termed encoding presets 
optimized for specific use cases and target devices based on empirical data, industry standards, and encoding techniques for different scenarios. Using presets, video encoders can efficiently handle the vast possible encoding configurations without specifying each parameter manually.
Presets directly impact the compression efficiency (\cf Fig.~\ref{fig:30fps_bitrate_psnr_plot} and \ref{fig:30fps_bitrate_VMAF_plot}). Moreover, encoding time and energy consumption increase exponentially for slower presets (\cf Fig.~\ref{fig:30fps_bitrate_time_plot} and \ref{fig:30fps_bitrate_energy_plot}). Faster presets typically employ simpler algorithms and fewer encoding passes, with reduced computations required for motion estimation, transform coding, and entropy coding. The result is a reduced overall computational effort and a lower energy consumption~\cite{Silveira2017Performance}. Faster presets also shorten the processing time, which saves energy by keeping the hardware components active for shorter durations.

\begin{figure*}[t]
\centering
\includegraphics[width=\linewidth]{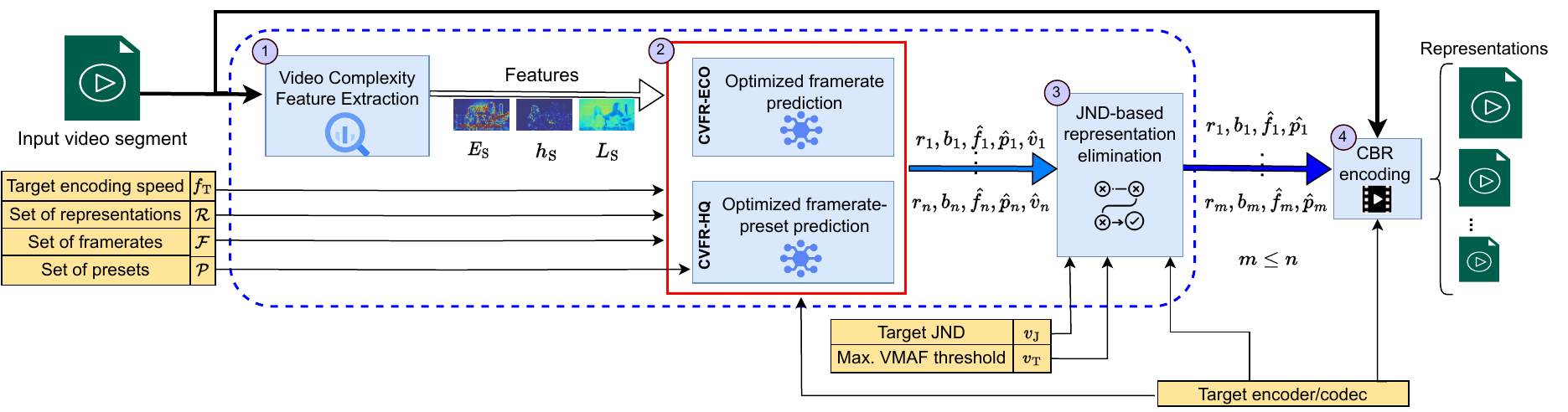}
\caption{Low-latency live encoding using \cvfr.}
\label{fig:gvfr_arch}
\end{figure*}

The x264~\cite{x264_ref} and x265~\cite{x265_ref} encoders offer ten pre-defined presets ranging from the highest compression efficiency and energy consumption (preset 9, known as \texttt{placebo}) to the fastest encoding speed and lowest energy consumption (preset 0, known as \texttt{ultrafast}). The encoder parameters depend on the particular encoding preset configuration~\cite{nasiri_multi_preset, caps_ref}. Generally, live content prefers the fastest encoding preset (\texttt{ultrafast}), independent of its dynamic complexity. Although this conservative technique achieves the intended result of live encoding with low encoding energy, the resulting visual quality is sub-optimal, especially for dynamically changing content~\cite{perf_ref}. Furthermore, when the content becomes easier to encode (\ie slow-motion videos, predictable frames with simple textures), the encoder achieves a higher speed than the target encoding speed. Configuring the preset to reduce this higher speed while still being compatible with the target live encoding speed improves the quality of the encoded content. However, when the content becomes complex again, the encoder preset must return to the faster configuration that achieves live encoding speed~\cite{sota_ref1}.

\subsection{Related variable preset encoding works}
Ramachandran~\etal~\cite{pradeep_ref} proposed an architecture using a proportional integral derivate (PID) module leveraged for content-adaptive live encoding. The proposed PID controller dynamically monitors the encoder's achieved framerate and adjusts parameters to maintain the framerate at the expected level while maximizing quality. However, the implementation details are proprietary and limited.

Nasiri~\etal~\cite{nasiri_multi_preset}  proposed an ML method to estimate bitrate ladders of multi-preset encoders for VoD applications. This method offers a reference content-adaptive bitrate ladder that exhaustively uses the fastest known preset of a given encoder. Then, an offline trained regressor transforms the fast-preset ladder into the slow-preset ladder needed for the encoding pass. However, the features used for prediction (\ie GLCM and temporal coherency) are computationally intensive and unsuitable for real-time live-streaming.

Our previous work \caps~\cite{caps_ref} determines the optimized preset for a given target bitrate representation to maintain the target encoding speed for low-latency encoding using XGBoost-based models~\cite{xgboost_ref}. This method improves the visual quality of representations at lower bitrates.

\subsection{Summary}
There are a few related works on variable preset encoding. Ramachandran~\etal~\cite{pradeep_ref} is proprietary, while Nasiri~\etal~\cite{nasiri_multi_preset} applies only to VoD applications. This paper considers two common approaches to low-latency encoding:
\paragraph{Low encoding energy}referred to as ecological \cvfreco mode, opts for the fastest preset while meeting encoding time constraints in scenarios where real-time or near-real-time encoding is critical.
\paragraph{Highest compression efficiency}referred to as high-quality \cvfrhq mode, maximizes the compression efficiency by encoding at slower presets and maintaining the encoding speed higher than the video framerate.

%% file: sec_proposed_method.tex
\section{Content-adaptive Variable Framerate (\cvfr) encoding scheme}
\label{sec:proposed}
Fig.~\ref{fig:gvfr_arch} presents the architecture of the proposed \cvfr scheme for live video streaming applications.
\cvfr receives the video segment initially and extracts its complexity features. \cvfr then offers two modes of operation:
\paragraph{Ecological} \cvfreco determines the optimized framerate for each bitrate ladder representation~\cite{DASH_Survey} using the fastest available preset.
\paragraph{High-quality} \cvfrhq predicts the representation's optimized framerate and preset using the video complexity features extracted for every segment and the set of pre-defined framerates and presets supported by the service provider.

The JND-based representation elimination algorithm ensures that the adjacent rate-distortion (RD) points of the bitrate ladder have a perceptual quality difference of at least one JND. Prediction for every segment is necessary because of the reasonably uniform frame-to-frame spatiotemporal content of the frames within a segment~\cite{DASH_Survey}. The JND-based representation elimination uses the average JND quality, assuming VMAF as the ``optimal'' measure of perceptual quality, predicted for each bitrate-resolution and framerate-preset configuration. 
The constant bitrate (CBR) encoding process uses the selected optimized bitrate-resolution and framerate-preset configurations. 

\begin{table}[t]
\centering
\caption{Notations used in \cvfr.}
\label{CVFR_notation}
\resizebox{\linewidth}{!}{
\begin{tabular}{c|l}
\specialrule{.12em}{.05em}{.05em}
\specialrule{.12em}{.05em}{.05em}
\emph{Notation} & \emph{Description}\\
\specialrule{.12em}{.05em}{.05em}
\specialrule{.12em}{.05em}{.05em}
\multicolumn{2}{c}{\emph{Video complexity features}}\\\hline
\ES & Average luma texture energy of segment\\
\hS & Average gradient of the luma texture energy of segment \\
\LS & Average luminescence of segment \\
\hline
\multicolumn{2}{c}{\emph{Input parameters}}\\\hline
$\mathcal{R}$ & Set of bitrate ladder representations\\
$\mathcal{F}$ & Set of supported framerates\\
$\mathcal{P}$ & Set of supported presets\\
$f_{\text{T}}$ & Target encoding speed [fps]\\
\vJ & Target JND\\
$v_{\text{T}}$ & Maximum VMAF threshold\\
\hline
\multirow{2}{*}{$r_{t},b_{t},\hat{f}_{t},\hat{p}_{t},\hat{v}_{t}$} & Resolution, bitrate and predicted framerate, preset and \\
& VMAF of the $t$\textsuperscript{th} representation\\
\specialrule{.12em}{.05em}{.05em}
\specialrule{.12em}{.05em}{.05em}
\end{tabular}}
\end{table}

Therefore, \cvfr encoding has four connected phases depicted in Fig.~\ref{fig:gvfr_arch}:
\begin{enumerate}[topsep=0pt,leftmargin=*,label=\protect\circled{\arabic*}]
  \item Video complexity feature extraction (Section~\ref{sec:feature_extraction});
  \item Optimized framerate and preset prediction (Section~\ref{sec:bitrate_ladder_pred});
  \item JND-based representation elimination (Section~\ref{sec:rep_elim});
  \item CBR encoding of the segment using the predicted bitrate-resolution and framerate-preset configurations.
\end{enumerate}
Table~\ref{CVFR_notation} summarizes the model notation for convenience.

\subsection{Video complexity feature extraction}
\label{sec:feature_extraction}
In video streaming applications, convolutional neural networks~\cite{3d_cnn_vqa_ref} is an intuitive method for feature extraction but cause inherent disadvantages for live-streaming applications~\cite{jtps_ref}, such as longer training time, inference time, and storage requirements. Although such methods can result in rich features, simpler models yielding significant prediction accuracy are more suitable for video streaming. Two popular state-of-the-art video complexity features are spatial information and temporal information~\cite{siti_itu_ref}, offering low feature correlation with the encoding output features such as bitrate and encoding time, which is insufficient for encoding parameter prediction in streaming applications~\cite{vca_ref}.

This paper uses three \DCT{}energy-based features, the average luma texture energy $E_{\text{Y}}$, the average gradient of luma texture energy $h$, and the average luminescence $L_{\text{Y}}$ extracted using the open-source \emph{video complexity analyzer (VCA)}, and use them as spatial and temporal complexity~\cite{vca_ref, jtps_ref}.   
These energy features averaged across all segment frames represent the average complexity segment characteristics \ES, \hS, and \LS.

\subsection{Optimized framerate and preset prediction}
\label{sec:bitrate_ladder_pred}
The optimized framerate and preset prediction method comprises two steps: modeling and optimization.

\subsubsection{Modeling} We define a set of video representation \mbox{$\mathcal{R}=\left\{\left({r}_i, {b}_i\right)| 1 \le i \le q\right\}$} for a pair of  encoding resolution $r_i$ and bitrate $b_i$.
The perceptual quality $v_{\left(r_{t},b_{t},f_{t},p_{t}\right)}$ and encoding speed $s_{\left(r_{t},b_{t},f_t,p_t\right)}$ of the $t$\textsuperscript{th} representation in $\mathcal{R}$ relies on video complexity features \ES, \hS, \LS, encoding resolution $r_t$, target bitrate $b_t$, framerate $f_t$, and preset $p_t$ parameters:
\begin{align}
\label{eq:v_pred}
    v_{\left(r_{t},b_{t},f_{t},p_{t}\right)} &= f_{V}\left(E_{\text{S}}, h_{\text{S}}, L_{\text{S}}, r_{t}, b_{t}, f_{t}, p_{t}\right);\\
\label{eq:s_pred}
    s_{\left(r_{t},b_{t},f_t,p_t\right)} &= f_{S}\left(E_{\text{S}}, h_{\text{S}}, L_{\text{S}}, r_{t}, b_{t}, f_{t}, p_{t}\right).
\end{align}
Spatiotemporal features encapsulate intricate spatial details and temporal dynamics within the video segment and help assess the video fidelity~\cite{csvt_ref1}. Including resolution, bitrate, framerate, and preset parameters in the models acknowledges the interplay between compression efficiency, temporal smoothness, and spatial clarity in shaping perceived quality. A higher resolution, bitrate, or framerate may improve the quality and increase the file size of the video segment. A slower preset at the same target bitrate can reduce the file size of the video segment. Similarly, a higher resolution, bitrate, framerate, or a slower preset can reduce the encoding speed.


\subsubsection{Optimization} of the framerate-preset pairs for each target representation utilizes the perceptual quality and encoding speed models developed in the previous step.

\paragraph{\cvfreco} involves a dual commitment to perceptual fidelity and encoding energy conservation. First, it chooses the fastest preset supported by the encoder $p_{1}$ for all representations to save energy. Second, it predicts the optimized framerate $\hat{f}_{t}$ from the supported framerates $\mathcal{F}$ for the $t$\textsuperscript{th} representation of $\mathcal{R}$ based on video complexity features \ES, \hS, \LS, and the target bitrate-resolution pair $\left(r_{t},b_{t}\right)$ maintaining the encoding speed above the threshold $f_{\text{T}}$;
\begin{align}
   \hat{f}_{t} &= \argmax_{f \in \mathcal{F}} \hat{v}_{\left(r_{t},b_{t},f,p_{1}\right)} & c.t. \hspace{1em} \hat{s}_{\left(r_{t},b_{t},f,p_{1}\right)} \geq f_{\text{T}};\\
   \hat{v}_{t} &= \max_{f \in \mathcal{F}} \hat{v}_{\left(r_{t},b_{t},f,p_{1}\right)} & c.t. \hspace{1em} \hat{s}_{(r_{t},b_{t},f,p_{1})} \geq f_{\text{T}}.  
\end{align}
where $\hat{v}_{(r_{t},b_{t},f,p_{1})}$ and $\hat{s}_{(r_{t},b_{t},f,p_{1})}$ are the predicted VMAF and encoding speed of the $t$\textsuperscript{th} representation in $\mathcal{R}$, using framerate $f$ and preset $p_{1}$. 
Thus, \cvfreco estimates the framerates synergizing the perceptual quality enhancement, energy reduction, and real-time processing imperatives. Therefore, the encoding configuration for the $t$\textsuperscript{th} representation in $\mathcal{R}$ is $\left(r_{t}, b_{t}, \hat{f}_{t}, p_{1}\right)$.

\paragraph{\cvfrhq} represents a strategic amalgamation of perceptual quality enhancement and computational efficiency. This scheme selects specific combinations of framerates and presets that maximize VMAF for each representation. By integrating temporal coherence, content complexity, and encoding efficiency, \cvfrhq optimizes the perceptual quality of encoded videos while adhering to real-time processing constraints. It jointly predicts the optimized framerate and preset of the $t$\textsuperscript{th} representation to maximize the compression efficiency while maintaining the encoding speed below the threshold  $f_{\text{T}}$. 
\cvfrhq ensures that the encoding speed exceeds the video framerate and maintains real-time feasibility. The optimization function is: 
\begin{align}
   (\hat{f}_{t},\hat{p}_{t}) &= \argmax_{f \in \mathcal{F} \land p \in \mathcal{P}} \hat{v}_{\left(r_{t},b_{t},f,p\right)} \hspace{1em} c.t. \hspace{1em} \hat{s}_{\left(r_{t},b_{t},f,p\right)} \geq f_{\text{T}};\\
   \hat{v}_{t} &= \max_{f \in \mathcal{F} \land p \in \mathcal{P}} \hat{v}_{\left(r_{t},b_{t},f,p\right)} \hspace{1em} c.t. \hspace{1em} \hat{s}_{\left(r_{t},b_{t},f,p\right)} \geq f_{\text{T}}. 
\end{align}
Thus, the encoding configuration for the $t$\textsuperscript{th} representation in $\mathcal{R}$ is $\left(r_{t}, b_{t}, \hat{f}_{t}, \hat{p}_{t}\right)$. 

\begin{algorithm}[t]
\caption{JND-based representation elimination.}
\small
\textbf{Input:}\\
\quad $q$~: number of representations in $\mathcal{R}$\\
\quad $\mathcal{R}=\bigcup_{t=1}^q\left\{\left(r_{t},b_{t},\hat{f}_{t},\hat{p}_{t}\right)\right\}$:  \begin{flushright}representations with predicted framerate and preset \end{flushright}
\quad $\hat{v}_{t}~; 1 \leq t \leq q$: predicted VMAF\\
\quad $v_{\text{T}}$~: maximum VMAF threshold \\
\quad $v_{J}$~: average target JND \\
\textbf{Output:}  $\hat{\mathcal{R}}=\left(r,b, \hat{f}, \hat{p}\right)$: set of  encoding configurations\\
    \begin{algorithmic}[1]
    \STATE $\hat{\mathcal{R}} \gets \left\{\left(r_{1},b_{1}, \hat{f}_{1}, \hat{p}_{1}\right)\right\}$ \label{alg:r_hat_set} \\
    \STATE $u \gets 1$  \label{alg:u_initialization}\\
    \IF{$\hat{v}_{1} \geq v_{\text{T}}$} \label{alg:vmaf_R1_comparison}
        \STATE \Return $\hat{\mathcal{R}}$  \label{alg:retR}
    \ENDIF
        \STATE $t \gets 2$  \label{alg:t_initialization}\\ 
        \WHILE{$t \leq q$}  \label{alg:while_start}
            \IF{$\hat{v}_{t} - \hat{v}_{u} \geq$ \vJ} \label{alg:vmaf_comparison}
                \STATE $\hat{\mathcal{R}} \gets \hat{\mathcal{R}}\cup \left\{\left(r_{t},b_{t}, \hat{f}_{t}, \hat{p}_{t}\right)\right\}$\\
                \STATE $u \gets t$  \label{alg:add_R_hat} \\
                \IF{$\hat{v}_{t} \geq v_{\text{T}}$}  \label{alg:vmaf_comp}
                    \STATE \Return $\hat{\mathcal{R}}$ \label{alg:loop_retR}
                \ENDIF
            \ENDIF
            \item $t \gets t + 1$
        \ENDWHILE    \label{alg:while_end}      
    \STATE \Return $\hat{\mathcal{R}}$
    \end{algorithmic}
\label{algo:res_eliminate}
\end{algorithm}

\subsection{JND-based representation elimination}
\label{sec:rep_elim}
In practice, the VMAF scores of consecutive representations are similar and introduce perceptual redundancy in the bitrate ladder. To address this issue, \cvfr leverages the minimum JND threshold required by the human eye to perceive differences in quality~\cite{lin2015experimental, wang2016mcl, wang2017videoset} since
perceptually redundant representations yield diminishes perceptual gains relative to their computational and energy costs. Eliminating redundant representations obviates associated encoding tasks, reducing processing time and energy consumption, and contributes to a more resource-efficient encoding workflow, conserving energy and reducing operational costs. While~\cite{zhu2022framework,jasla_ref} explored complex VMAF-based JND thresholds unsuitable for live-streaming applications, \cvfr adopts a fixed JND threshold $v_{\text{J}}$ as input from the streaming service provider.

\begin{table*}[!t]
\caption{Experimental parameters of \cvfr used in this paper.}
\centering
\begin{tabular}{l|c|c|c|c|c|c|c|c|c|c|c}
\specialrule{.12em}{.05em}{.05em}
\specialrule{.12em}{.05em}{.05em}
\multicolumn{2}{c|}{\emph{Parameter}} &  \emph{Symbol} & \multicolumn{9}{c}{\emph{Values}}\\
\specialrule{.12em}{.05em}{.05em}
\specialrule{.12em}{.05em}{.05em}
\multirow{2}{*}{\emph{Set of representations}} & \emph{Resolution height [pixels]} & \multirow{2}{*}{$\mathcal{R}$} & 234 & 360 & 432 & 432 & 540 & 720 & 720 & 1080 & 1080 \\
& \emph{Bitrate [Mbps]} &  & 0.145 & 0.365 & 0.730 & 1.100 & 2.000 & 3.000 & 4.500 & 6.000 & 7.800 \\
\hline
\multicolumn{2}{c|}{\emph{Set of framerates [fps]}} & $\mathcal{F}$ & \multicolumn{9}{c}{7.5, 15, 24, 30}\\
\hline
\multicolumn{2}{c|}{\emph{Set of presets [x264]}} & $\mathcal{P}$ & \multicolumn{9}{c}{0 (\texttt{ultrafast}) -- 8 (\texttt{veryslow})}\\
\hline
\multicolumn{2}{c|}{\emph{Encoding speed threshold [fps]}} & $f_{\text{T}}$ & \multicolumn{9}{c}{30}\\
\hline
\multicolumn{2}{c|}{\emph{Average target JND}} & \vJ & \multicolumn{3}{c}{2} & \multicolumn{3}{|c}{4} & \multicolumn{3}{|c}{6}\\
\hline
\multicolumn{2}{c|}{\emph{Maximum VMAF threshold}} & $v_{\text{T}}$ & \multicolumn{3}{c}{98} & \multicolumn{3}{|c}{96} & \multicolumn{3}{|c}{94}\\
\specialrule{.12em}{.05em}{.05em}
\specialrule{.12em}{.05em}{.05em}
\end{tabular}
\label{tab:exp_par}
\end{table*}

Algorithm~\ref{algo:res_eliminate} implements the JND-based representation elimination. This algorithm receives the number of representations $q$ in $\mathcal{R}$ comprising their predicted framerate and preset ${(r_{t},b_{t},\hat{f}_{t},\hat{p}_{t})}$ and VMAF $\hat{v}_{t}$ (from Section~\ref{sec:bitrate_ladder_pred}),  maximum VMAF threshold $v_{\text{T}}$, and average target JND \vJ. The first representation of $\mathcal{R}$ is always part of the encoding representation set $\hat{\mathcal{R}}$ (line~\ref{alg:r_hat_set}). 
When the predicted VMAF of the first representation in $\mathcal{R}$, \ie $\hat{v}_{1}$ is greater than $v_{\text{T}}$ (above which the representation is perceptually lossless), it eliminates all other representations from the bitrate ladder(lines~\ref{alg:vmaf_R1_comparison}--\ref{alg:retR}). 
If the predicted VMAF difference between the current representation $\hat{v}_{t}$ and the previously selected representation in $\hat{\mathcal{R}}$ is greater than (or equal to) \vJ, $\hat{\mathcal{R}}$ includes the current representation (lines~\ref{alg:vmaf_comparison}--\ref{alg:add_R_hat}). 
The algorithm terminates when the predicted VMAF of the current representation is higher than $v_{\text{T}}$ (lines~\ref{alg:vmaf_comp}--\ref{alg:loop_retR}). The algorithm loops between lines~\ref{alg:while_start} and~\ref{alg:while_end} until it analyzes all representations in $\mathcal{R}$.  Finally, $\hat{\mathcal{R}}$ is the algorithm's output.

%% file: sec_experimental_design.tex
\section{Experimental Design}
\label{sec:exp_design}
This section illustrates our experimental design to assess the performance of \cvfr. We evaluate \cvfr, comparing it with five distinct benchmark schemes regarding energy consumption and compression efficiency.

\subsection{Experimental setup}
\label{sec:test_methodology}

We run experiments on a dual server with Intel Xeon Gold 5218R processors (80 cores, operating at 2.10 GHz). We execute the VCA and x264 encoder using eight threads, using x86 SIMD optimizations~\cite{x86_simd_ref}. We use the video complexity dataset (VCD)~\cite{VCD_ref} consisting of five hundred 2160p resolution video sequences (segments) and consider the x264 version r3107~\cite{x264_ref} as the target encoder. We adopt the bitrate-resolution pairs specified in the Apple HLS authoring specifications~\cite{HLS_ladder_ref} as the set of representations $\mathcal{R}$. 
We extract the spatiotemporal features, \ES, \hS, and \LS, using VCA v2.0~\cite{vca_ref} to assess video complexity. We select the average target JND $v_{\text{J}}$ according to current industry practices~\cite{kah_ref,jnd_streaming} and set the maximum VMAF threshold $v_{\text{T}}$ accordingly, \ie $v_{\text{T}} = 100 - v_{\text{J}}$. 
Table~\ref{tab:exp_par} summarizes the list of experimental parameters.

\subsection{Prediction models}
\label{sec:imp_pred}
To ensure the robustness and generalization of the prediction models, we perform a five-fold cross-validation scheme for video sequences and average the results. The scheme also ensures splitting the test and training segments. Our approach to designing prediction models is scalable by training models for each preset supported by the streaming service provider, avoiding retraining the entire network when adding a new preset. This work uses three prediction models for comparing their accuracy in terms of the coefficient of determination ($R^{2}$) and mean absolute error (MAE): \textit{(i)} linear regression~\cite{lr_ref}, \textit{(ii)} XGBoost~\cite{xgboost_ref} and \textit{(iii)} random forest regression~\cite{rf_ref}.

\begin{table}[t]
\caption{Prediction accuracy of VMAF and encoding speed prediction models for \texttt{ultrafast} preset for VCD dataset~\cite{VCD_ref} using x264 AVC encoder.}
\centering
\begin{tabular}{l|c|c|c|c}
\specialrule{.12em}{.05em}{.05em}
\specialrule{.12em}{.05em}{.05em}
& \multicolumn{2}{c|}{VMAF prediction} & \multicolumn{2}{c}{Encoding speed prediction} \\
Method & $R^{2}$ & MAE & $R^{2}$ & MAE \\
\specialrule{.12em}{.05em}{.05em}
\specialrule{.12em}{.05em}{.05em}
Linear regression & 0.748 & 7.988 & 0.716 & 252.744 \\
XGBoost &  0.882 &  4.873  & 0.946 &  51.325  \\
\textbf{Random forest} & \textbf{0.895} &  \textbf{4.552} & \textbf{0.949} &  \textbf{39.200} \\
\specialrule{.12em}{.05em}{.05em} 
\specialrule{.12em}{.05em}{.05em}
\end{tabular}
\label{tab:vmaf_pred_models_res}
\end{table}

Table~\ref{tab:vmaf_pred_models_res} shows the results of the VMAF prediction and the encoding speed of the \texttt{ultrafast} preset using the default hyperparameters of the models. We observe that the $R^{2}$ score is the maximum, and MAE is the minimum for random forest models. Therefore, we use random forest for VMAF and encoding speed prediction for each encoding preset in our experiments. We perform the hyperparameter tuning on the prediction models on the \texttt{ultrafast} preset to balance the size and prediction accuracy of the models. The selected hyperparameters~\cite{scikit-learn} are as follows:
\begin{enumerate}
    \item minimum number of samples required to be at a leaf node (\texttt{min\_samples\_leaf}) set to 1;
    \item minimum number of samples required to split an internal node (\texttt{min\_samples\_split}) set to 2;
    \item number of trees in forest (\texttt{n\_estimators}) set to 100;
    \item maximum depth of the tree (\texttt{max\_depth}) set to 14.
\end{enumerate}

\subsection{Benchmarks}
We compare \cvfr with five schemes carefully selected according to the following paragraphs:
\paragraph{Default} adopts a fixed framerate of 30 fps and the \texttt{ultrafast} preset for the CBR encoding of the HLS bitrate ladder~\cite{HLS_ladder_ref}.
\paragraph{Bruteforce-ECO} \cite{netflix_paper} determines the optimized framerate for each representation by encoding all framerates with the \texttt{ultrafast} preset, representing the ideal bitrate ladder constructed using \cvfreco, \ie when the prediction models are \SI{100}{\percent} accurate.  
\paragraph{Bruteforce-HQ} \cite{netflix_paper} determines the optimized framerate-preset pair for each representation by encoding all framerates and presets, indicating the ideal bitrate ladder constructed using \cvfrhq, \ie when the prediction models are \SI{100}{\percent} accurate. 
\paragraph{Herrou~\etal}~\cite{vfrc_ref}  determine the lowest framerate that does not affect the perceived original video quality. Combining two successive binary random forest classifiers uses 32 features, including the pixel luminance map, frame difference magnitude, and horizontal and vertical coordinates of motion vectors.
\cite{vfrc_ref} is the only work applied to live-streaming essential for our evaluation, as presented in Table~\ref{tab:vfr_sota}. 

\paragraph{\caps}~\cite{caps_ref} determines the optimized preset for each representation for a fixed framerate of 30\,fps. 


\subsection{Evaluation metrics}
We evaluate the \cvfr encoding using the following metrics:
\begin{description}[font=\normalfont\itshape]
\item[\BDRP{} and \BDRV]\cite{DCC_BJDelta} refer to the average increase in bitrate of the representations compared to the reference bitrate ladder encoding scheme to maintain the same PSNR and VMAF. A negative BDR suggests a boost in the coding efficiency of the considered encoding scheme compared to the reference bitrate ladder encoding scheme.
\item[BD-PSNR and BD-VMAF] refer to the average increase in PSNR and VMAF at the same bitrate compared with the reference bitrate ladder encoding scheme. Positive \mbox{BD-PSNR} and BD-VMAF denote an increase in the coding efficiency of the considered encoding scheme compared to the reference bitrate ladder encoding.
\item[Relative storage space difference] between the considered encoding scheme $b_{\text{opt}}$ and the reference encoding scheme $b_{\text{ref}}$ to store all bitrate ladder representations: 
\begin{equation}
    \Delta S = \frac{\sum b_{\text{opt}}}{\sum b_{\text{ref}}} - 1.
\end{equation}
\item[Encoding energy consumption] measures the CPU energy consumption on Linux using the Running Average Power Limit (RAPL) interface and the \texttt{CodeCarbon} tool~\cite{codecarbon_ref}.
\item[Storage energy consumption] of all server data~\cite{bianco2016energy}:
\begin{equation}
E_{\text{sto}} = S_{\text{d}} \cdot P_{\text{b}} \cdot T_{\text{s}},
\label{eq:storageEnergy}
\end{equation}
where $S_{\text{d}}$ is the video data size in \si{\bit}, $P_{\text{b}}$ is the power consumption per bit in \si{\watt\per\bit}, and $T_{\text{s}}$ is the time duration taken for data to be stored in \si{\hour}. In our experiments, we measure $T_{\text{s}} = \SI{1.9}{\giga\byte\per\second}$ using the Unix \texttt{dd} command and set $P_{\text{b}} = 7.84 \cdot \SI{e-12}{\watt\per\bit}$~\cite{bianco2016energy}. 
\end{description}

%% file: sec_evaluation.tex
\section{Experimental Results}
\label{sec:evaluation}
We conduct three experiments to evaluate \cvfr in four areas:
\begin{enumerate*}[label=\roman*)]
\item prediction models,
\item coding efficiency analysis, 
\item storage consumption analysis, and
\item energy consumption.
\end{enumerate*}

\subsection{Prediction models}
In this section, we \textit{a)} assess the accuracy of our VMAF and encoding speed prediction models, \textit{b)} explore the relative importance of features, \textit{c)} analyze average framerate-preset predictions, and \textit{d)} discuss latency considerations.

\begin{figure}[t]
\centering
\begin{subfigure}{0.48\columnwidth}
    \centering
    \includegraphics[clip,width=\textwidth]{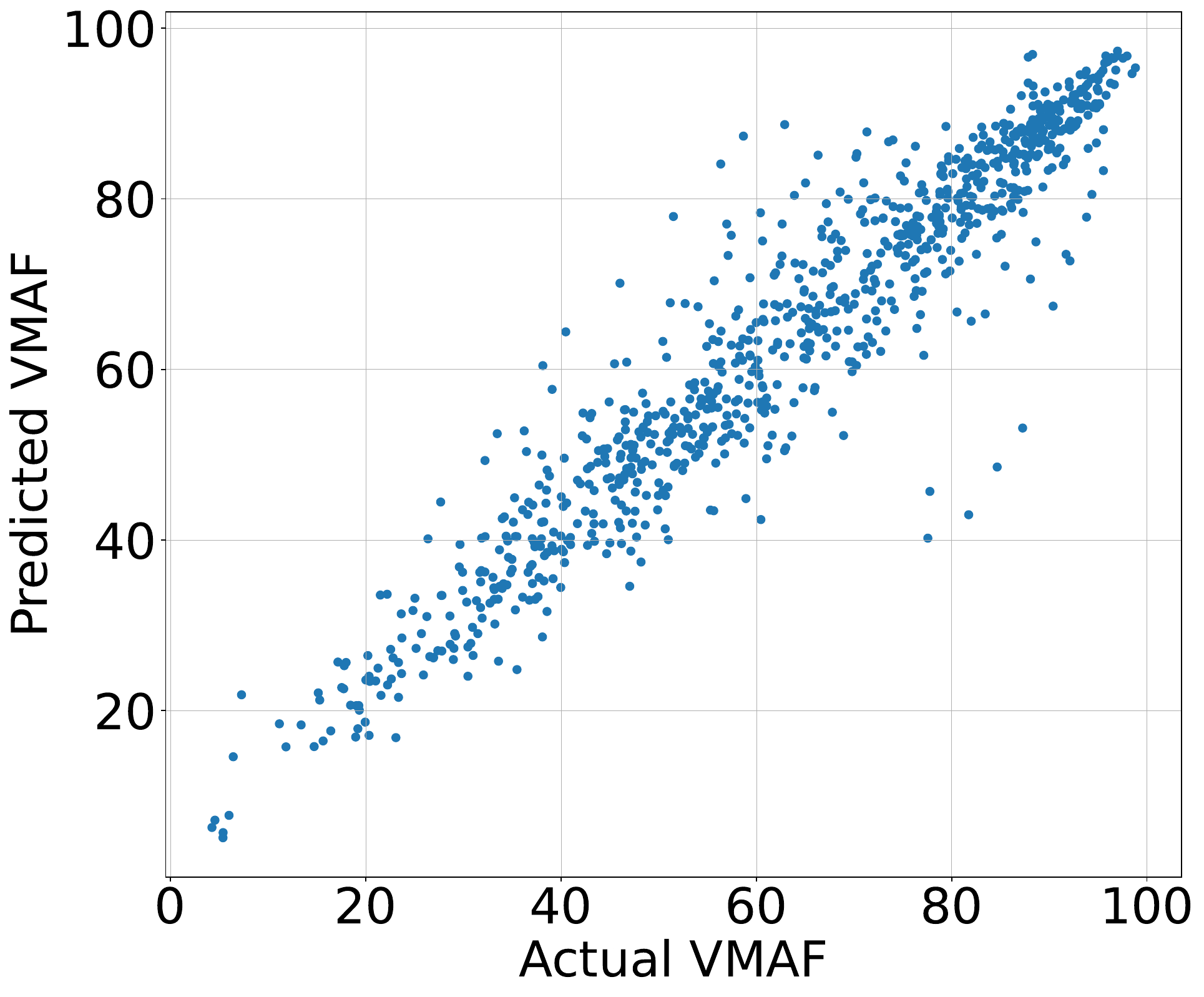}
    \caption{VMAF.}
    \label{fig:vmaf_pred_scatter}
\end{subfigure}
\hfill
\begin{subfigure}{0.50\columnwidth}
    \centering
    \includegraphics[clip,width=\textwidth]{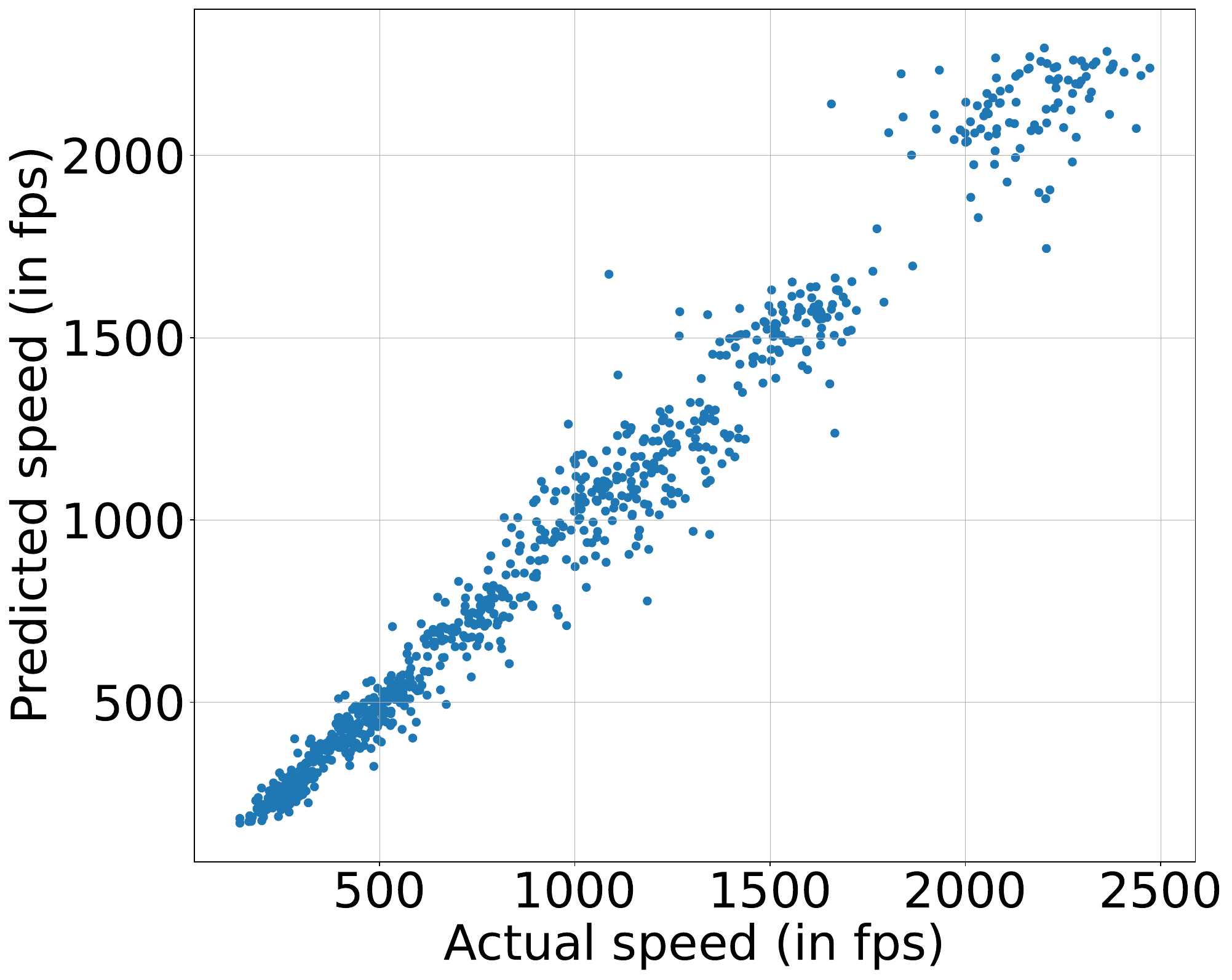}
    \caption{Encoding speed.}
    \label{fig:speed_pred_scatter}    
\end{subfigure}
\hfill
\begin{subfigure}{0.49\columnwidth}
    \centering
    \includegraphics[clip,width=\textwidth]{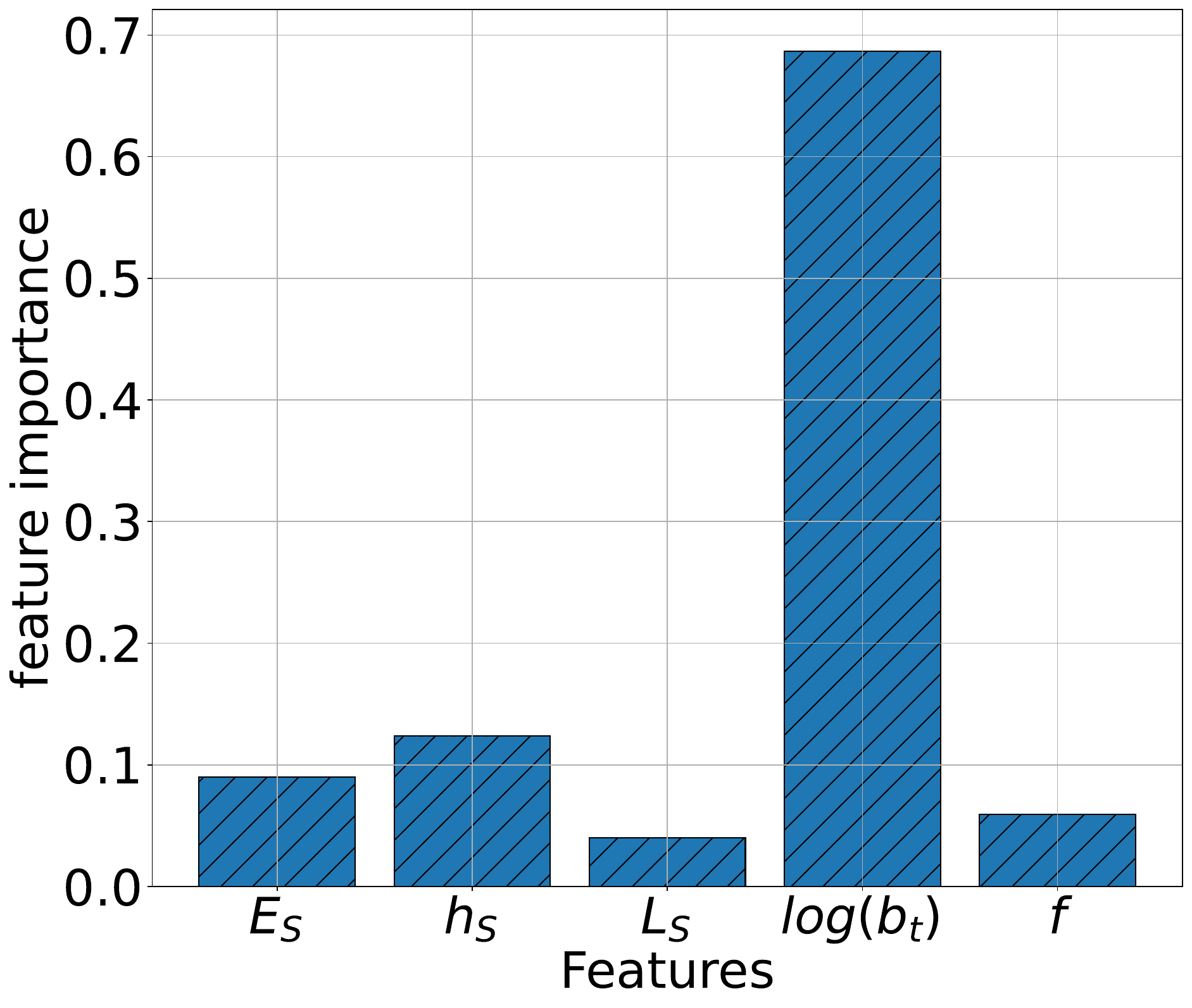}
    \caption{Feature relevance for VMAF prediction.}
    \label{fig:vmaf_pred_shap}
\end{subfigure}
\hfill
\begin{subfigure}{0.49\columnwidth}
    \centering
    \includegraphics[clip,width=\textwidth]{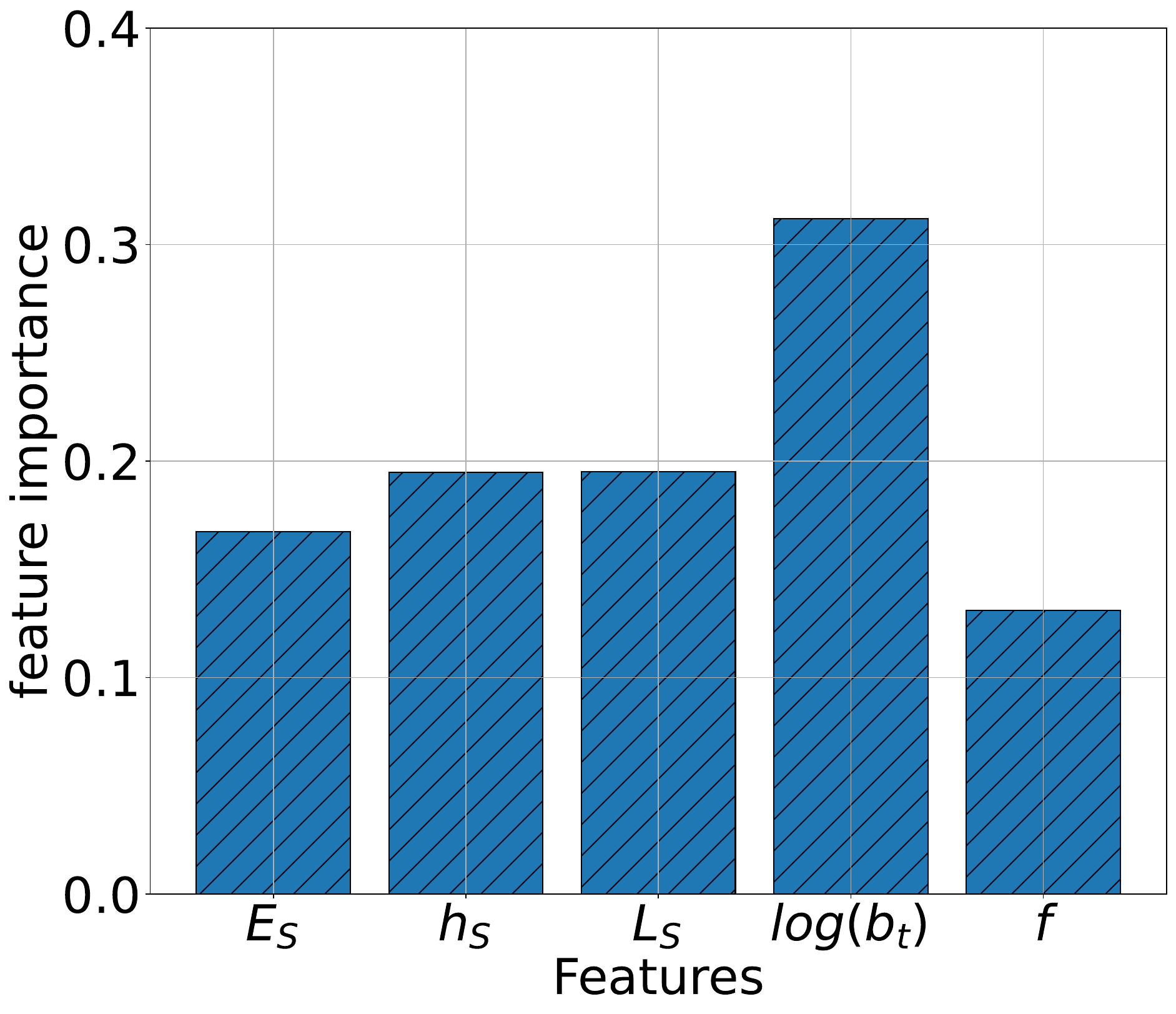}
    \caption{Feature relevance for encoding speed prediction.}
    \label{fig:speed_pred_shap}
\end{subfigure}
\begin{subfigure}{0.485\columnwidth}
    \centering
    \includegraphics[clip,width=\textwidth]{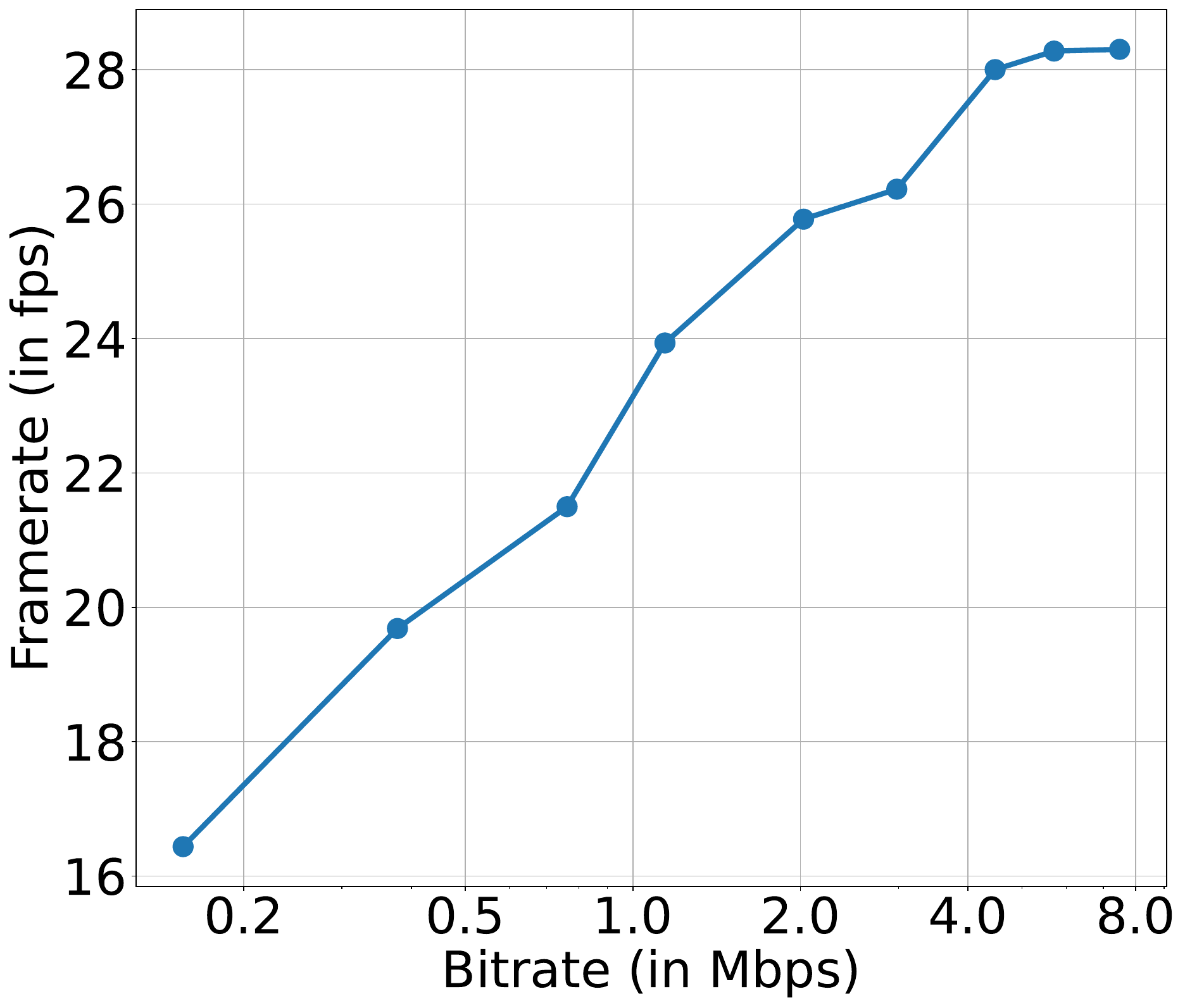}
    \caption{Average \cvfreco framerate.}
    \label{fig:avg_fr_plot_cvfreco}
\end{subfigure}
\hfill
\begin{subfigure}{0.495\columnwidth}
    \centering
    \includegraphics[clip,width=\textwidth]{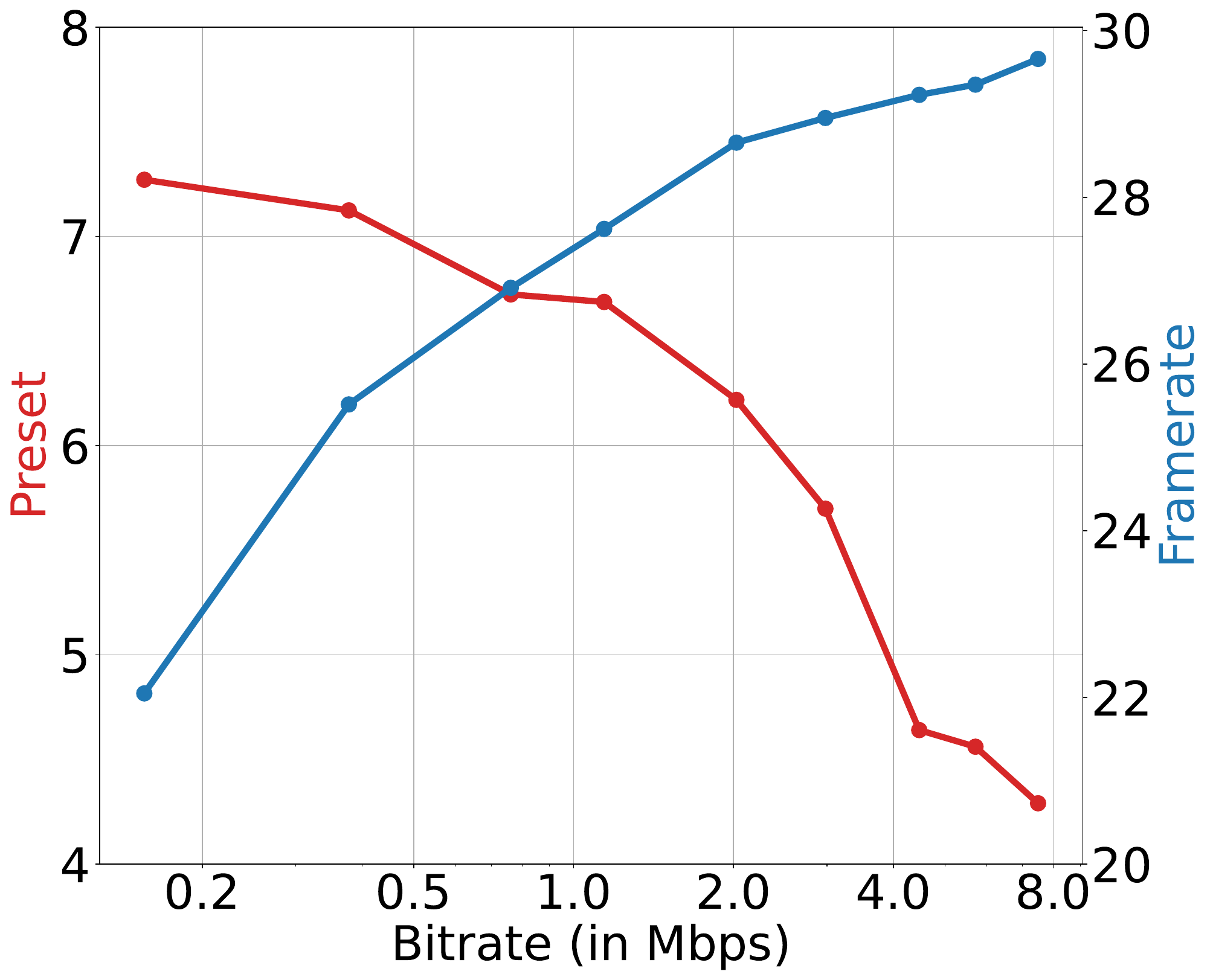}
    \caption{Average \cvfrhq framerate and preset.}
    \label{fig:avg_fr_preset_plot_cvfrhq}
\end{subfigure}
\caption{Prediction results.}
\end{figure}

\begin{figure*}[t]
\centering
\begin{subfigure}{0.245\textwidth}
    \centering
    \includegraphics[clip,width=\textwidth]{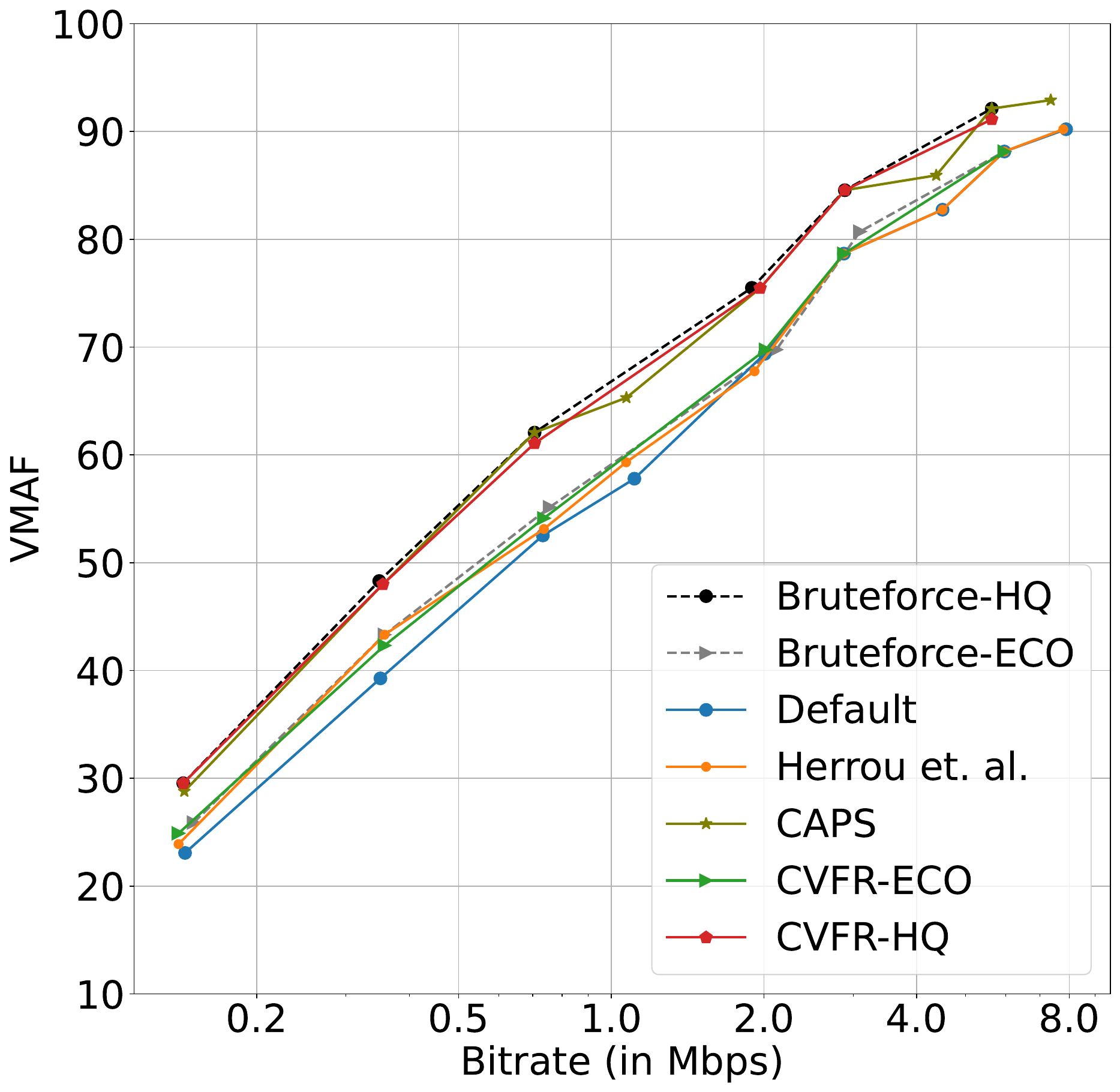}
    \caption{\texttt{Bunny\_s000}}
    \label{fig:bunny_rd}
\end{subfigure}
\hfill
\begin{subfigure}{0.245\textwidth}
    \centering
    \includegraphics[clip,width=\textwidth]{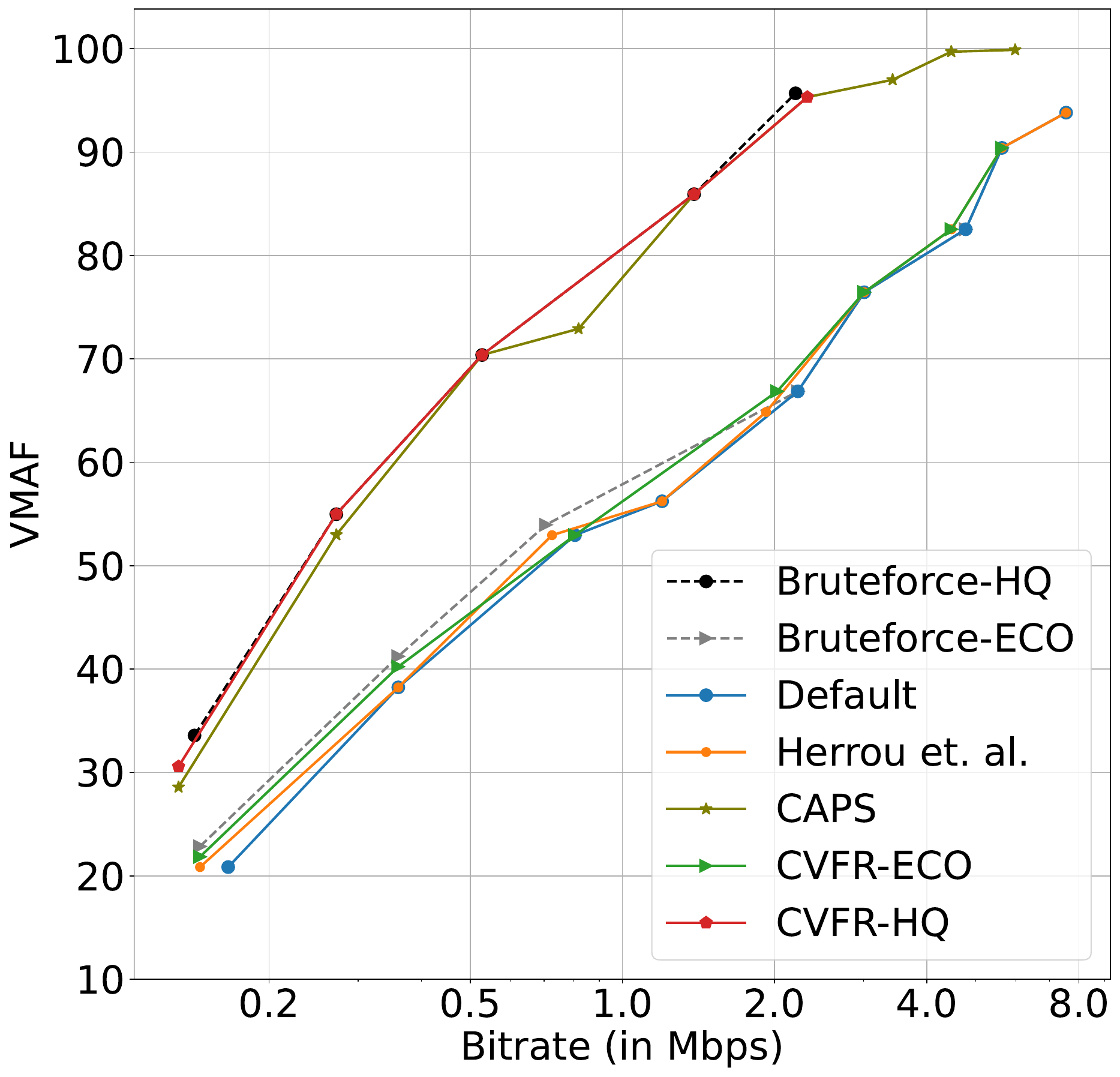}
    \caption{\texttt{Characters\_s000}}
\end{subfigure}
\hfill
\begin{subfigure}{0.245\textwidth}
    \centering
    \includegraphics[clip,width=\textwidth]{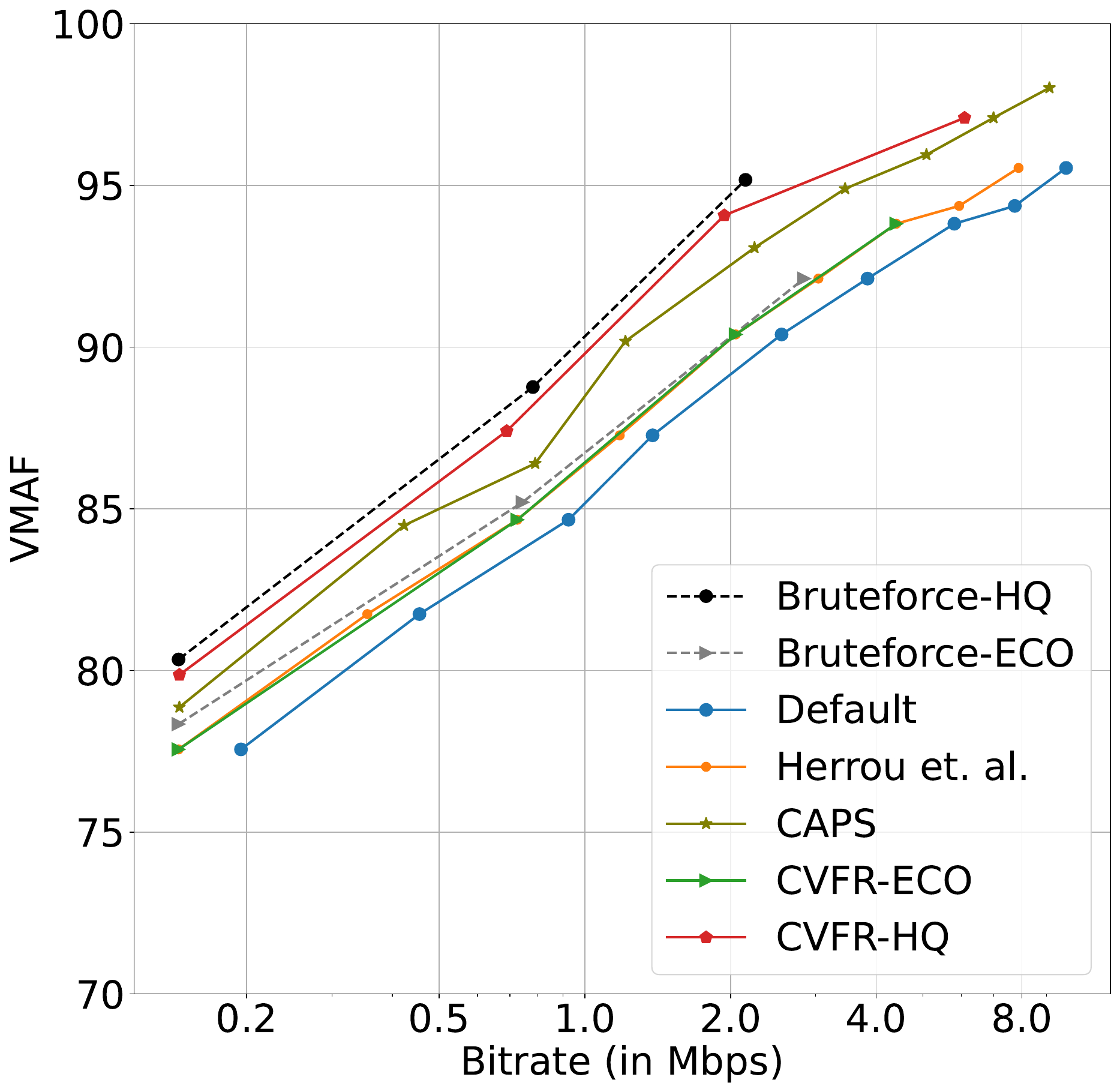}
    \caption{\texttt{Eldorado\_s000}}
\end{subfigure}
\hfill
\begin{subfigure}{0.245\textwidth}
    \centering
    \includegraphics[clip,width=\textwidth]{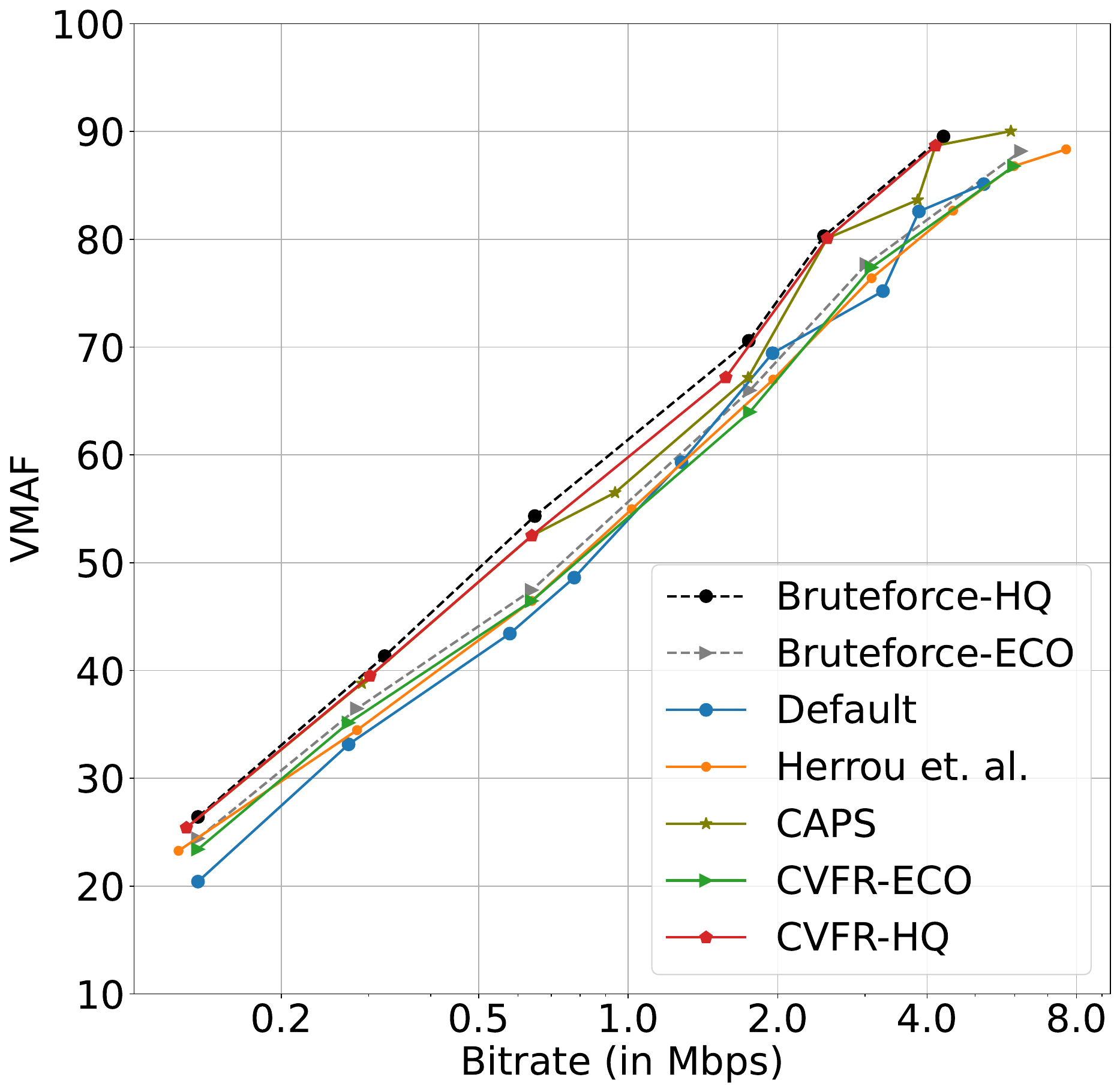}
    \caption{\texttt{Eldorado\_s005}}
\end{subfigure}
\vfill
\begin{subfigure}{0.245\textwidth}
    \centering
    \includegraphics[clip,width=\textwidth]{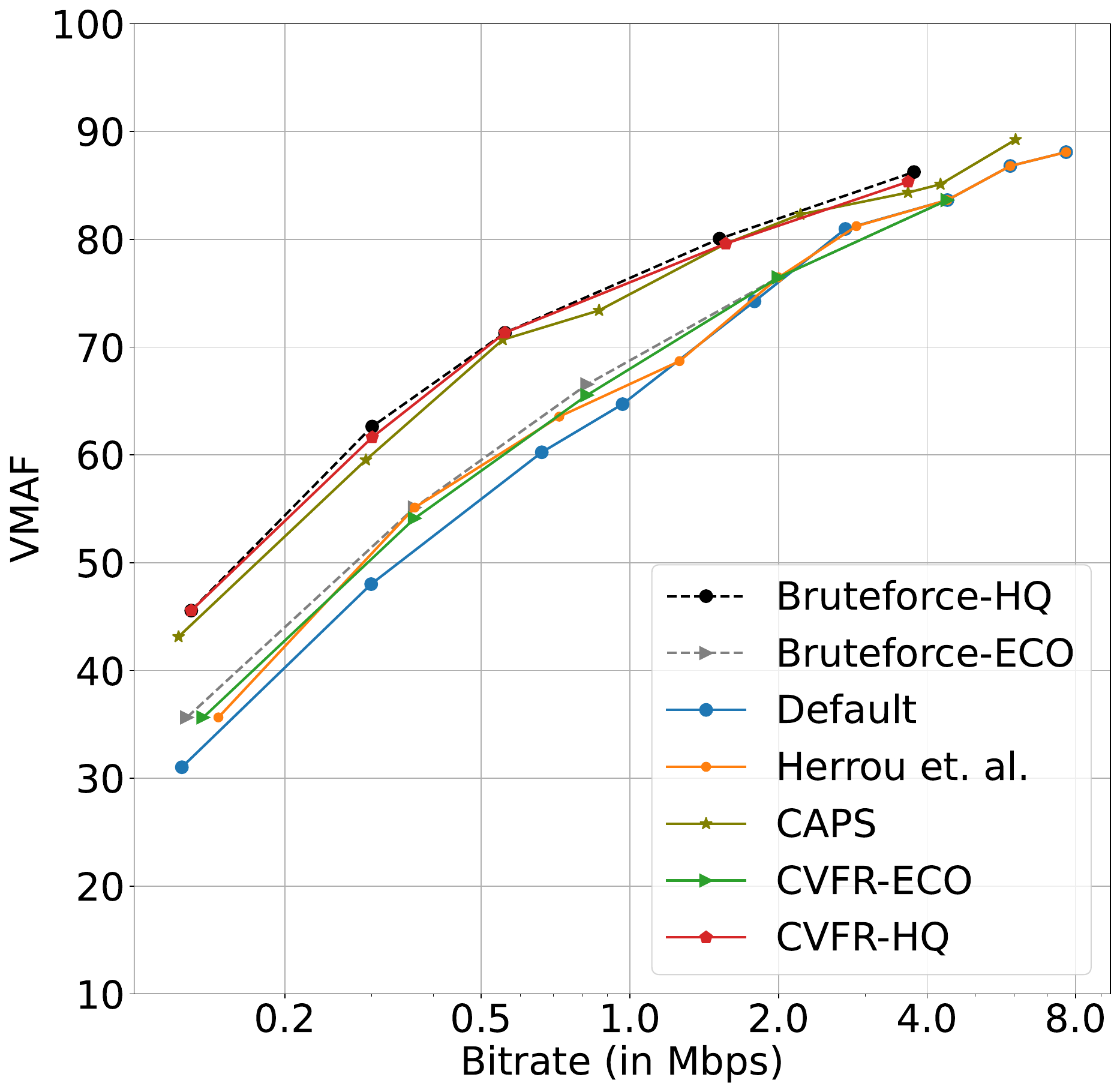}
    \caption{\texttt{HoneyBee\_s000}}
\end{subfigure}
\hfill
\begin{subfigure}{0.245\textwidth}
    \centering
    \includegraphics[clip,width=\textwidth]{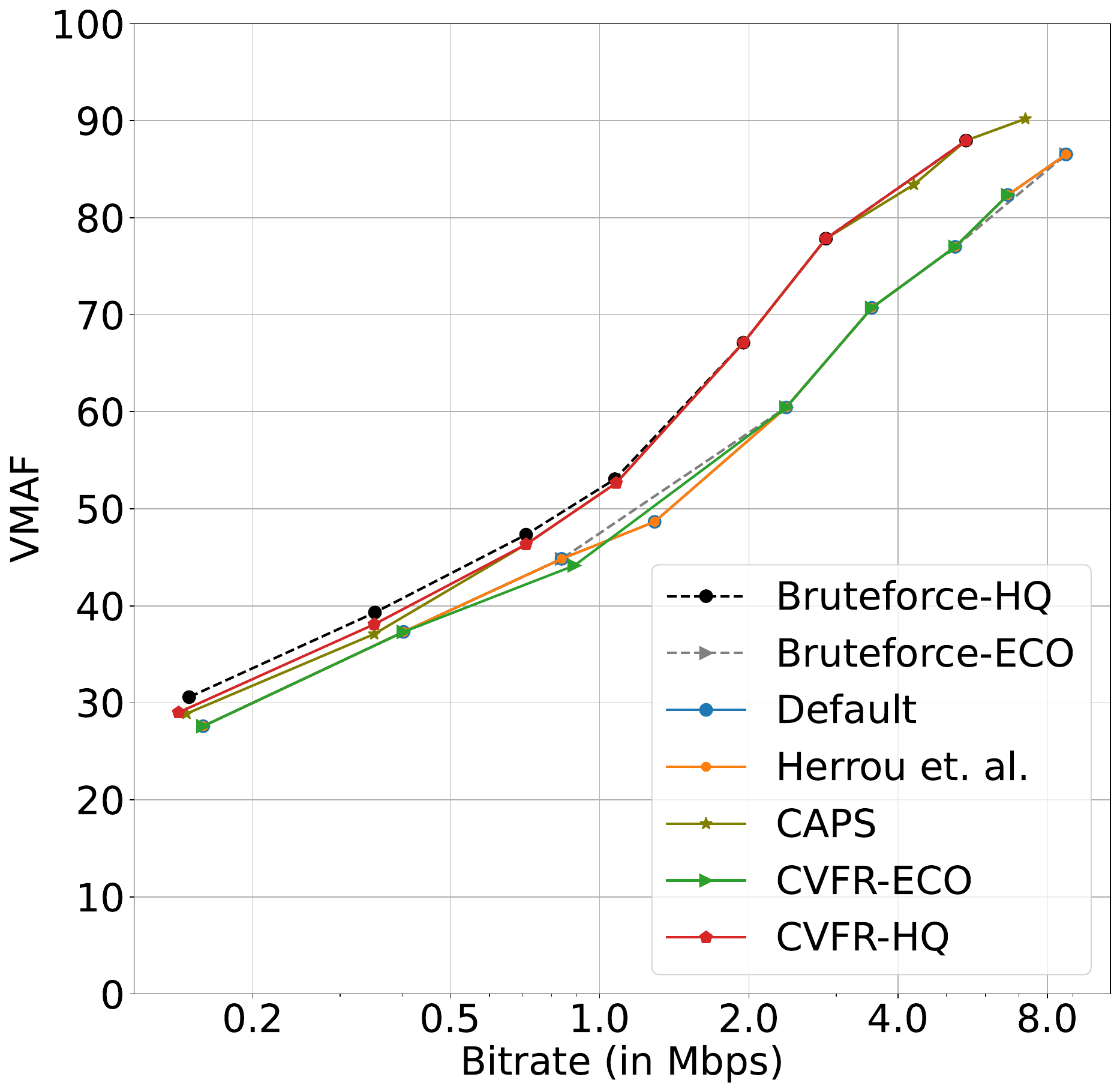}
    \caption{\texttt{RushHour\_s000}}
\end{subfigure}
\hfill
\begin{subfigure}{0.245\textwidth}
    \centering
    \includegraphics[clip,width=\textwidth]{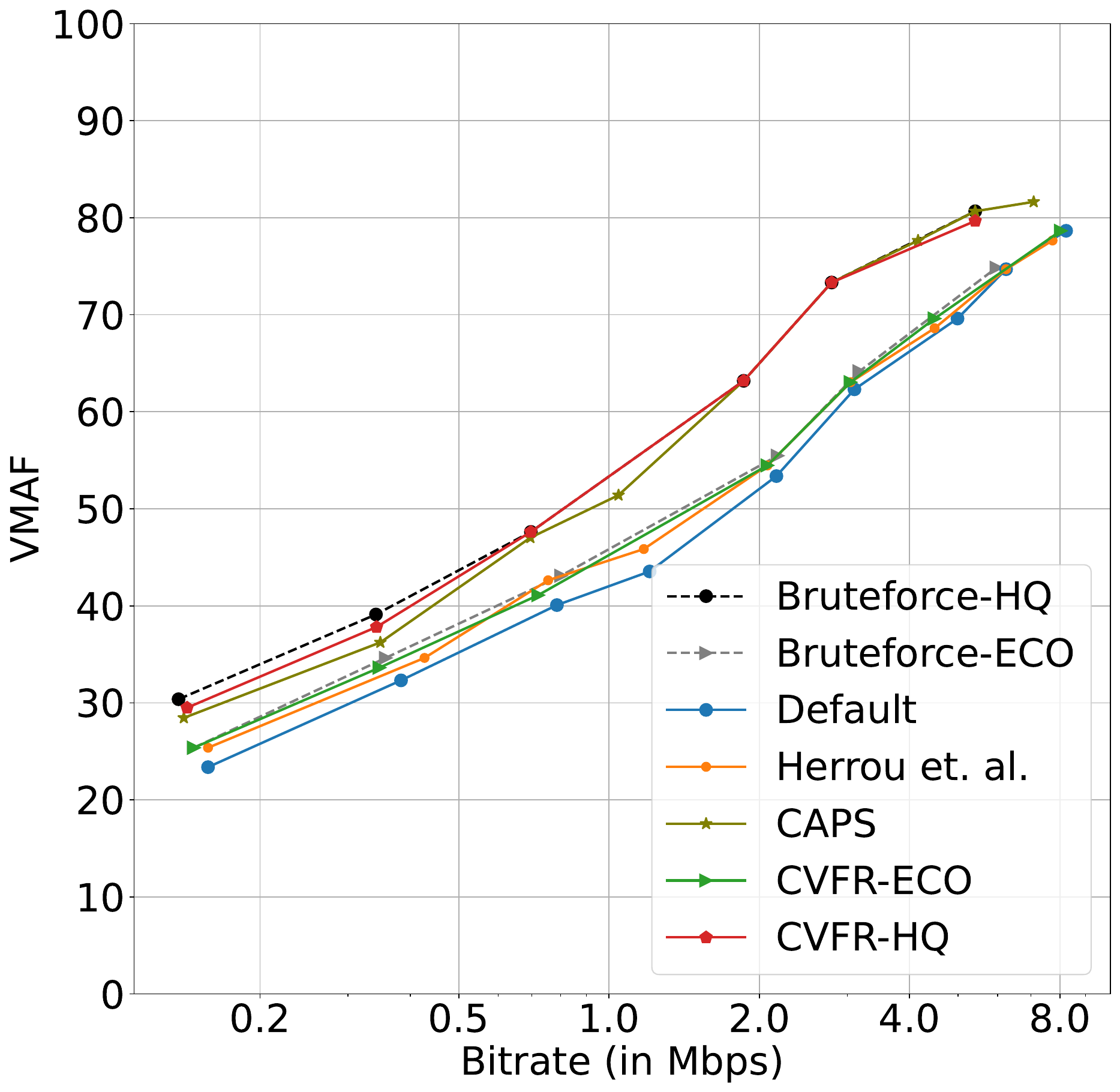}
    \caption{\texttt{Runners\_s000}}
\end{subfigure}
\hfill
\begin{subfigure}{0.245\textwidth}
    \centering
    \includegraphics[clip,width=\textwidth]{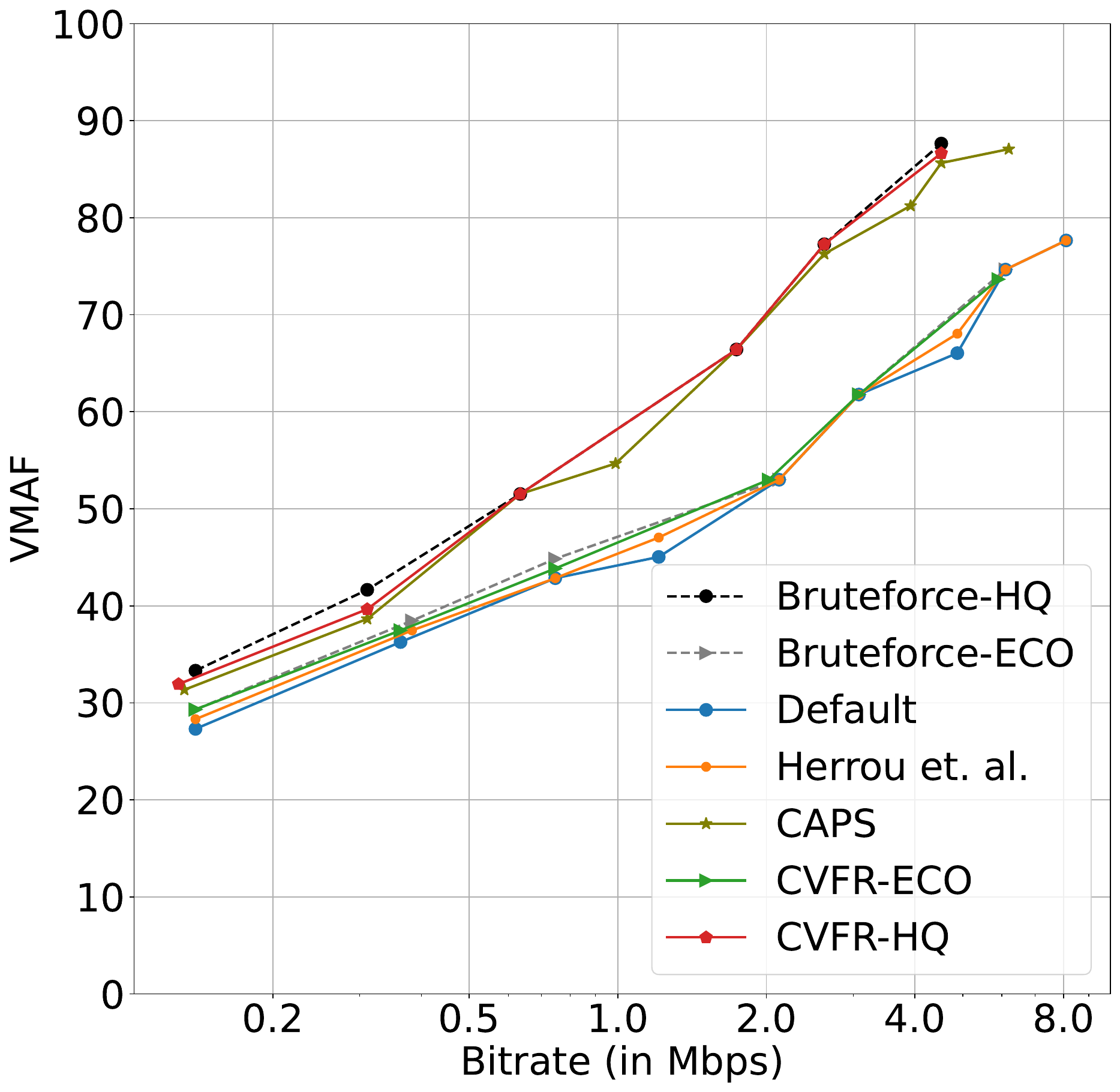}
    \caption{\texttt{Wood\_s000}}
\end{subfigure}
\caption{RD curves of representative video sequences (segments) in Table~\ref{tab:vidseq}.}
\label{fig:rd_res_gvfr}
\end{figure*}

\paragraph{Accuracy} We evaluate the accuracy of our VMAF and encoding speed prediction models against the ground truth values recorded within the training dataset (\cf Fig.~\ref{fig:vmaf_pred_scatter} and Fig.~\ref{fig:speed_pred_scatter}). The analysis reveals an average $R^{2}$ score of  0.886 for VMAF and 0.968 for encoding speed prediction models, showing a strong positive correlation of the prediction. Furthermore, the average MAE of the VMAF and encoding speed prediction models are 4.765 and 45.848, respectively. Noteworthy, the minor differences in VMAF scores may not be perceptible or bothersome to the audience. Viewers are generally more concerned about smooth and uninterrupted streaming than small quality fluctuations. Hence, the prediction errors are acceptable for live-streaming applications.


\paragraph{Relative feature importance} Impurity-based feature importance measures the contribution of each input feature to accurate predictions within the ensemble~\cite{feature_imp_ref}, critical for model interpretability in random forest regression.
In this paper, we measure impurity using MAE. Features consistently leading to the most significant impurity reduction across all the forest decision trees are the most important. Fig.~\ref{fig:vmaf_pred_shap} and Fig.~\ref{fig:speed_pred_shap} show the impurity-based feature importance measures corresponding to the features utilized in the VMAF and encoding speed prediction models. The target bitrate in the logarithmic scale ($\log\left(b_{t}\right)$) is the most influential feature for VMAF prediction, followed by the \hS, \ES, framerate, and \LS. Similarly, the order of importance in encoding speed prediction is $\log\left(b_{t}\right)$, \hS, \LS, \ES, and framerate.


\paragraph{Framerate-preset prediction} We analyze the average framerate-preset predictions of \cvfr in Figs.~\ref{fig:avg_fr_plot_cvfreco} and~\ref{fig:avg_fr_preset_plot_cvfrhq}. On average, \cvfreco chooses 15 fps at 0.145\,Mbps and 30\,fps at 7.8\,Mbps, as it always selects the fastest preset for encoding (\texttt{ultrafast} in x264) as discussed in Section~\ref{sec:bitrate_ladder_pred}. Like \cvfreco, \cvfrhq selects a lower framerate when the target bitrate decreases. However, it tends towards slower presets as the target bitrate drops because slower preset encodings at lower bitrates can yield higher VMAF while satisfying the encoding speed constraint for low-latency encoding.


\paragraph{Latency} We evaluate the pre-processing latency ($\tau_{\text{p}}$) in encoding introduced by the video complexity feature extraction and the model inference to predict the optimized bitrate-resolution-framerate-preset configurations. We extract the features at an average rate of 352\,fps over the entire dataset (2160p resolution). This result is critical in future-proofing the system by handling evolving content requirements (\eg 8K resolution or high framerate content). The time to predict the framerate-preset for each representation is \SI{5}{\milli\second}. As video complexity feature extraction and the optimized framerate and preset prediction can execute concurrently in real applications, the overall latency introduced by \cvfr is negligible.

\subsection{Coding efficiency analysis}
\subsubsection{Rate-distortion (RD) analysis} Fig.~\ref{fig:rd_res_gvfr} analyzes the RD curves of \texttt{default} encoding, bruteforce encoding~\cite{netflix_paper}, \cvfreco encoding, and \cvfrhq encoding for selected video sequences (segments) of various video content complexities displayed in Table~\ref{tab:vidseq}. In most cases, \cvfrhq yields higher VMAF than the other encoding schemes at the same target bitrates for all video complexity classes because it is specifically optimized to maximize VMAF within the bounds of the target encoding speed. Moreover, the RD curve of \cvfrhq is very close to Bruteforce-HQ encoding~\cite{netflix_paper}, demonstrating the high accuracy of the VMAF and encoding speed prediction. \cvfreco yields higher VMAF than the default scheme at low bitrates due to the selection of lower framerates. Furthermore, the VMAF difference between consecutive RD points of \cvfreco and \cvfrhq is at least the target JND, assumed as six VMAF points in the figure.

\begin{table}[t]
    \centering
    \caption{Representative video sequences.}
    \begin{tabular}{l|r|r|r}
\specialrule{.12em}{.05em}{.05em}
\specialrule{.12em}{.05em}{.05em}
        Sequence & \ES & \hS & \LS \\
\specialrule{.12em}{.05em}{.05em}
\specialrule{.12em}{.05em}{.05em}
        \texttt{Bunny\_s000} & 22.40 & 4.70 & 129.21\\
        \texttt{Characters\_s000} & 45.42 & 36.88 & 134.56\\
        \texttt{Eldorado\_s000} & 15.28 & 49.76 & 140.54\\
        \texttt{Eldorado\_s005} & 100.37 & 9.23 & 109.06\\
        \texttt{HoneyBee\_s000} & 42.93 & 7.91 & 103.00\\
        \texttt{RushHour\_s000} & 47.75 & 19.70 & 101.66\\
        \texttt{Runners\_s000} & 105.85 & 22.48 & 126.60\\
        \texttt{Wood\_s000} & 124.72 & 47.03 & 119.57\\
\specialrule{.12em}{.05em}{.05em}
\specialrule{.12em}{.05em}{.05em}
    \end{tabular}
    \label{tab:vidseq}
\end{table}

\subsubsection{Bjøntegaard delta rates (BDR)}
We further evaluate the coding efficiency using \BDRP, \BDRV, BD-PSNR, and \mbox{BD-VMAF} compared to the default encoding, as shown in Table~\ref{tab:res_cons_energy}. Bruteforce encoding~\cite{netflix_paper} (with and without JND-based representation elimination) yields \SI{100}{\percent} accurate results representing the highest bound of the compression efficiency improvement (in VMAF) compared to the default encoding.

\begin{table}[t]
\caption{Average encoding performance compared to the \texttt{default} encoding.}
\centering
\resizebox{\columnwidth}{!}{
\begin{tabular}{@{}l@{ }||@{ }c@{ }|@{ }c@{ }|@{ }c@{ }|@{ }c@{ }|@{ }c@{ }|@{ }c@{ }|@{ }c@{ }|@{ }c@{}}
\specialrule{.12em}{.05em}{.05em}
\specialrule{.12em}{.05em}{.05em}
Method & \vJ & \BDRP & \BDRV & BD-PSNR & BD-VMAF & $\Delta S$  &  $\Delta E_{\text{enc}}$ & $\Delta E_{\text{sto}}$\\
& & & & [dB] & & &\\
\specialrule{.12em}{.05em}{.05em}
\specialrule{.12em}{.05em}{.05em}
Bruteforce-ECO &              & -17.81\%  & -16.41\%  & 0.73 & 5.94 & -10.94\%  &  267.11\% & -20.69\%  \\
Bruteforce-ECO & 2 & -19.61\%  & -18.11\%  & 0.78 & 6.66 & -23.93\%  &  267.11\% & -22.43\%  \\
Bruteforce-ECO & 4 & -20.13\%  & -19.00\% &  0.79 & 6.86 & -42.79\%  &  267.11\% &  -61.30\% \\
Bruteforce-ECO & 6 &  -20.15\% & -19.13\% & 0.79  & 6.89 & -54.59\%  &  267.11\% &  -75.59\%\\
\hline
Bruteforce-HQ &              & -52.33\%  & -37.90\%  & 2.39 & 9.89  & -10.94\%  &  19174.15\% & -20.69\%  \\
Bruteforce-HQ & 2 & -51.96\%  & -39.18\%  & 2.39 & 10.33 & -38.80\%  &  19174.15\% & -62.55\%  \\
Bruteforce-HQ & 4 & -52.56\%  & -40.04\%  & 2.44 & 10.62 & -52.87\%  &  19174.15\% &  -77.78\% \\
Bruteforce-HQ & 6 & -53.06\%  & -40.41\%  & 2.44 & 10.74 & -59.01\%  &  19174.15\% &  -83.20\%\\
\hline
Herrou~\etal~\cite{vfrc_ref} & & -8.75\% & -8.81\% & 0.38 &  3.90  & -1.35\%  & -1.29\%  & -2.72\% \\
\caps~\cite{caps_ref}  & & -43.24\%  & -30.46\%   & 2.12 &  8.57  & -10.73\% & 746.70\% & -20.37\% \\
\hline
\cvfreco &               & -16.37\%  &  -11.17\%  & 0.52 &  3.72  &   -1.18\%  & -10.33\%  & -20.38\%\\
\cvfreco & 2 & -18.38\%  &  -12.05\%  & 0.51 &  4.23  & -21.70\%  & -23.12\%  & -38.70\%\\
\cvfreco & 4  & -17.25\%  &  -13.22\%  & 0.53 &  4.43  & -40.15\%  & -39.54\%  & -64.18\%\\
\cvfreco & 6  & -17.91\%  &  -13.54\%  & 0.52 &  4.50  & -51.26\%  & \textbf{-48.64\%}  & -76.24\%\\
\hline
\cvfrhq &                & -51.69\%  & -37.44\%   & 2.28 & 9.13  & -21.11\%  & 638.84\%  &  -40.99\%\\
\cvfrhq & 2   & -53.42\%  & -38.81\%   & 2.38 & 9.85   & -44.41\%  & 426.15\%  &  -69.10\% \\
\cvfrhq & 4   & -54.00\%  & -39.53\%   & 2.42 & 10.09  & -55.83\%  & 322.03\%  &  -80.49\% \\
\cvfrhq & 6   & \textbf{-54.25\%}  & \textbf{-39.63\%}   & \textbf{2.43} & \textbf{10.14}  & \textbf{-58.98\%}  & 282.94\%  &  \textbf{-83.18\%}\\
\specialrule{.12em}{.05em}{.05em}
\specialrule{.12em}{.05em}{.05em}
\end{tabular}}
\label{tab:res_cons_energy}
\end{table}

\begin{figure}[t]
\centering
    \includegraphics[width=0.45\textwidth]{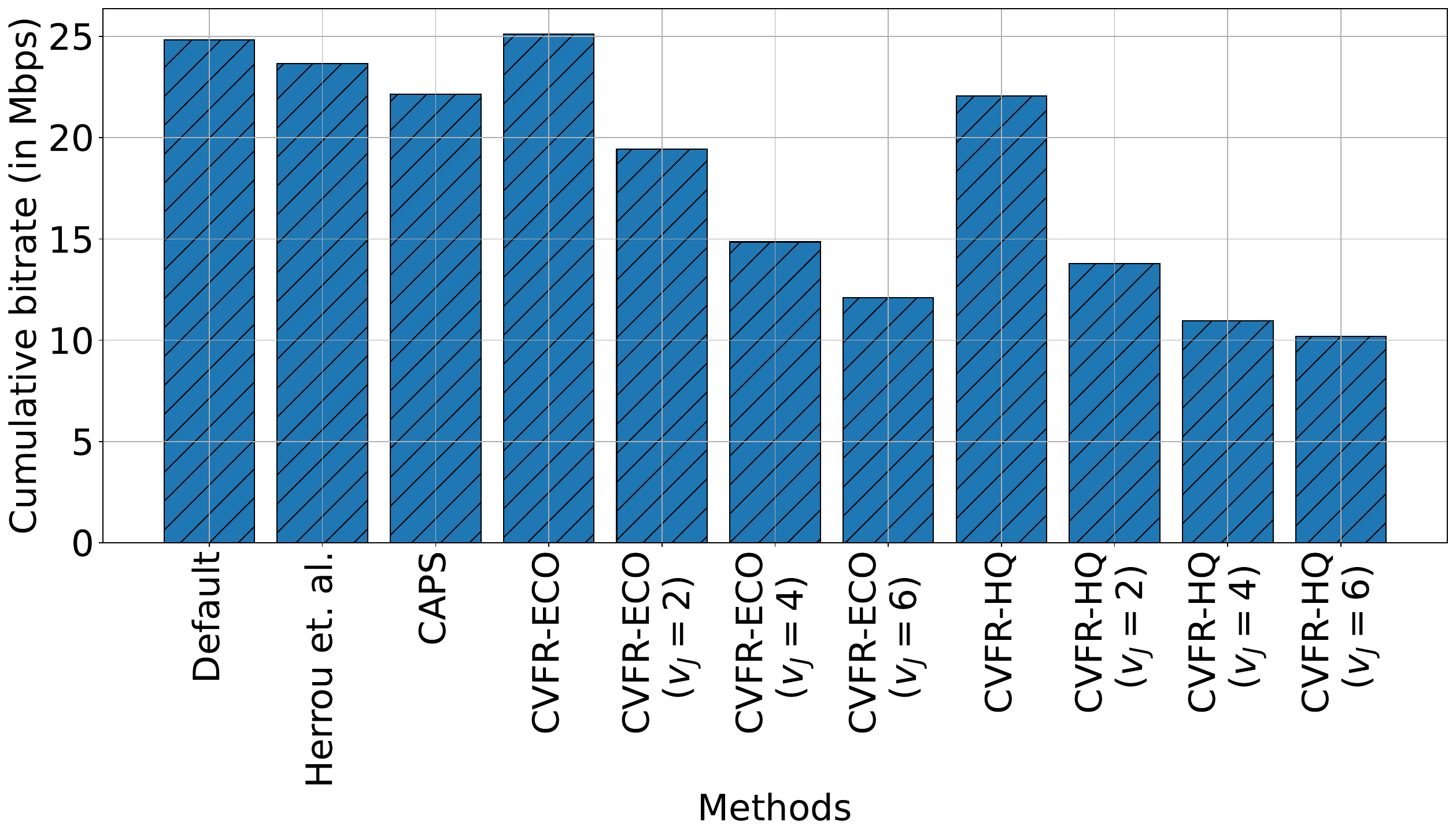}
    \caption{Average cumulative bitrate of each video segment using the considered encoding schemes.}
    \label{fig:storage_res}
\end{figure}

\begin{figure}[t]
\centering
\begin{subfigure}{0.450\textwidth}
    \centering
    \includegraphics[width=\textwidth]{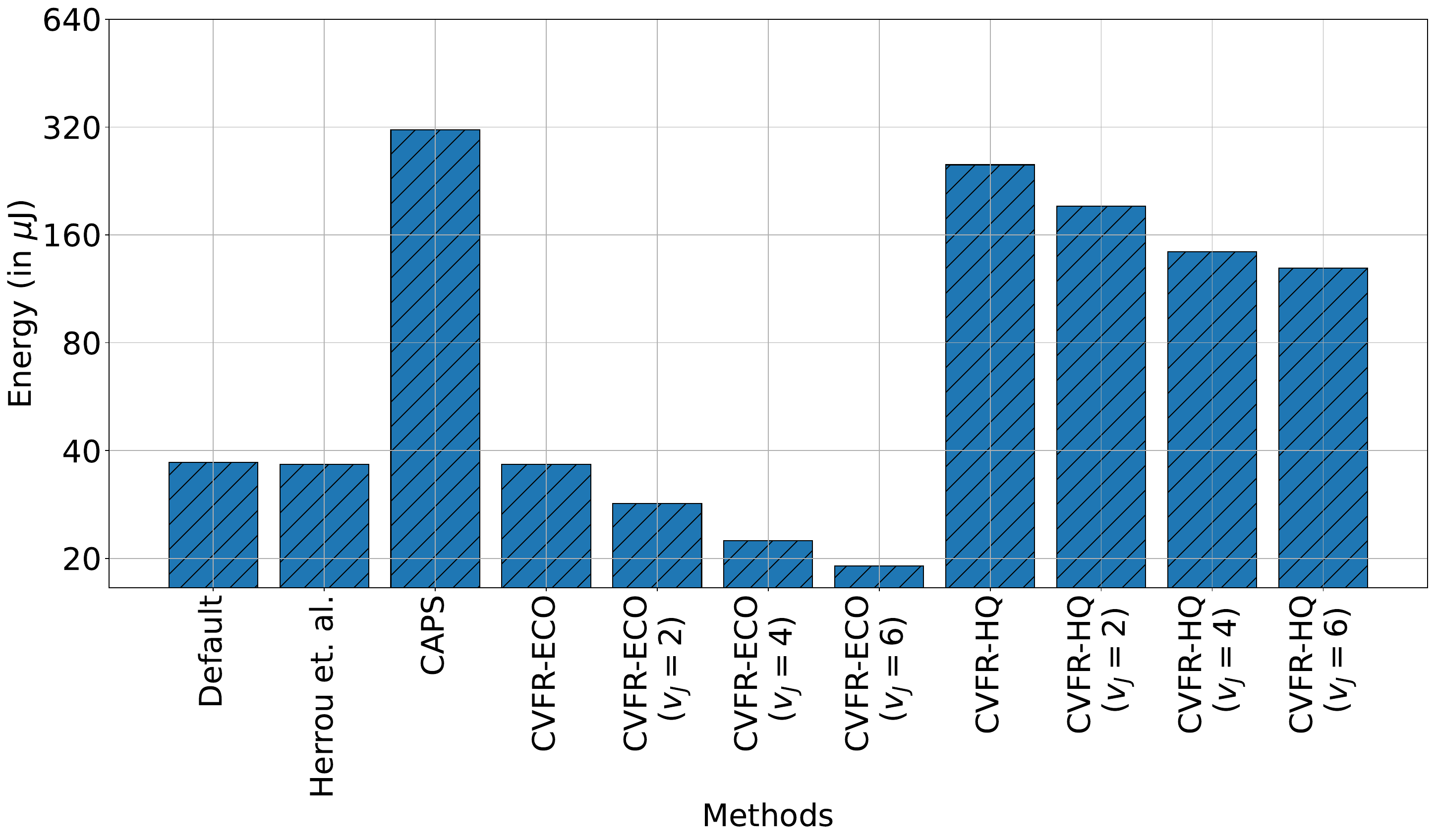}
    \caption{Encoding energy.}
    \label{fig:enc_energy_res}
\end{subfigure}
\hfill
\begin{subfigure}{0.450\textwidth}
    \centering
    \includegraphics[width=\textwidth]{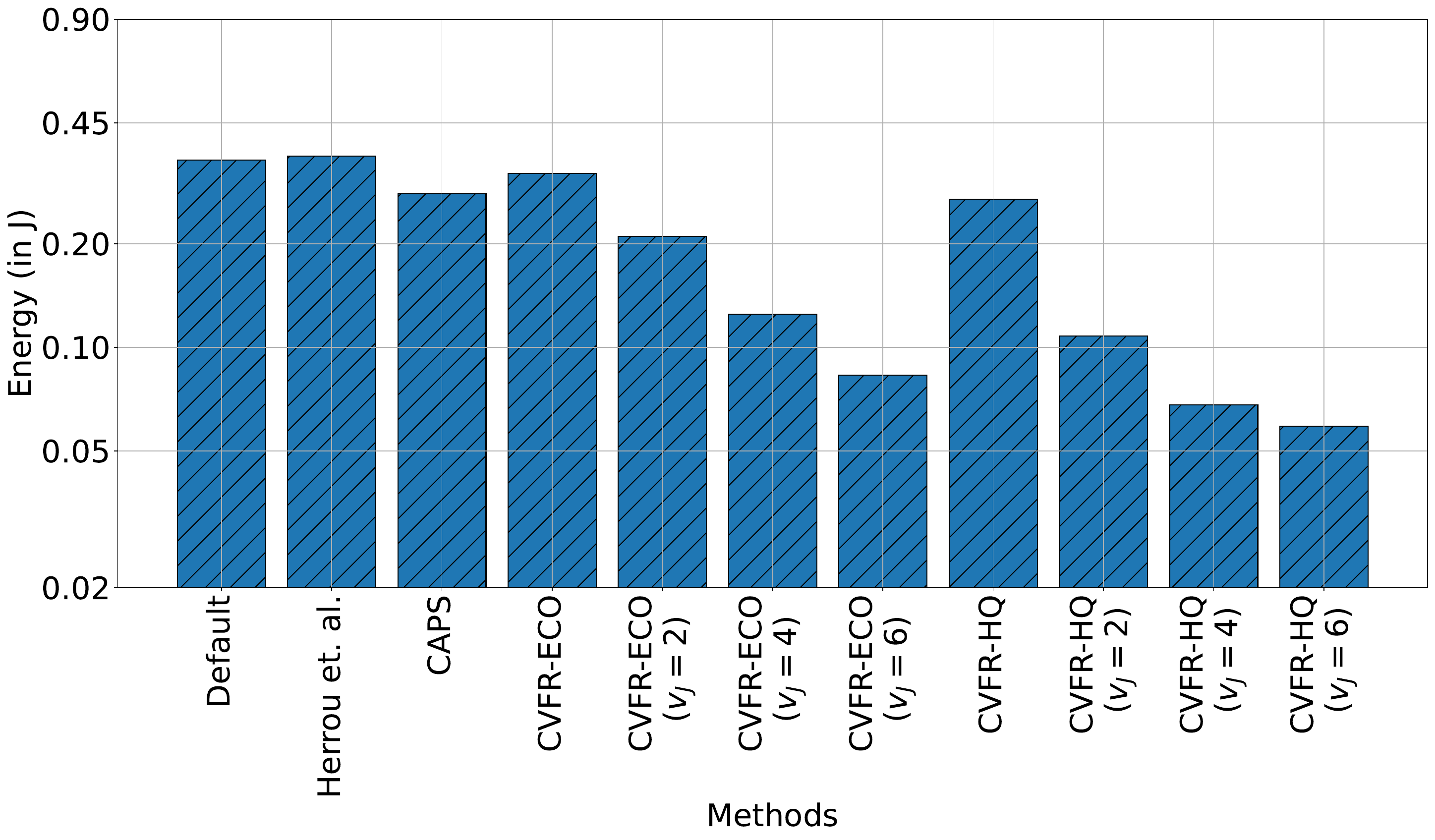}
    \caption{Storage energy.}
    \label{fig:storage_energy_res}
\end{subfigure}
\caption{Comparison of average encoding and storage energy consumption for each \SI{4}{\second} video segment using the considered encoding schemes.}
\end{figure}

\subsection{Storage consumption}
We evaluate the relative difference in the storage space between the considered encoding schemes and the default encoding scheme to store all bitrate ladder representations. 
\cvfrhq achieved a remarkable reduction in data size during video encoding (up to \SI{58.98}{\percent}) compared to the alternative schemes, as observed in Table~\ref{tab:res_cons_energy}, and ensures more efficient utilization of network resources. Higher $\Delta S$ translates to reduced storage requirements and lower delivery costs, which can be significant for large-scale streaming platforms. Lower cumulative bitrates place less strain on the network infrastructure, reducing the risk of network congestion.

\subsection{Energy consumption}
We conduct a comprehensive evaluation of encoding schemes by analyzing the relative differences in energy consumption during encoding ($\Delta E_{\text{enc}}$) and storage ($\Delta E_{\text{sto}}$) of the bitrate ladder compared to the reference encoding scheme.

\subsubsection{Encoding energy consumption}
Predictably, \cvfreco outperforms the other schemes in encoding energy, as shown in Fig.~\ref{fig:enc_energy_res}. Table~\ref{tab:res_cons_energy} shows a reduction in encoding energy consumption reduction for \cvfreco compared to the \texttt{default} encoding by \SI{10.33}{\percent}, which scales further up to \SI{48.64}{\percent}, using JND-based representation elimination (\vJ=6). \cvfreco encodes video using \texttt{ultrafast} preset, which yields the lowest encoding time and, subsequently, the lowest encoding energy consumption, in contrast to \caps and \cvfrhq. Moreover, lower framerates at low bitrates further reduce energy consumption. 
\caps results in the highest encoding energy, owing to the choice of slower presets at lower target bitrates. \cvfrhq yields lower energy consumption than \caps, attributed to the optimized framerate selection and preset selection. On average, \caps and \cvfrhq consume \SI{746.70}{\percent} and \SI{638.84}{\percent} more energy than the default encoding, respectively. However, the encoding energy of \cvfrhq is only  \SI{282.94}{\percent} higher than the default encoding, using JND-based representation elimination (\vJ=6).

\subsubsection{Storage energy consumption}
Fig.~\ref{fig:storage_energy_res} shows that \cvfrhq and all the other state-of-the-art methods yield reductions in storage energy compared to the \texttt{default} encoding, directly influenced by the size and the time required to store the data on the disk, as described in Eq.~\ref{eq:storageEnergy}. Consequently, despite its emphasis on maximizing coding efficiency and eliminating perceptually redundant representations, \cvfrhq achieves significant reductions in data size during video encoding, resulting in remarkable storage energy savings and, hence, streaming/transmission savings. According to Table~\ref{tab:res_cons_energy}, across various JND values, \cvfrhq demonstrates a noteworthy range of reductions, ranging from \SI{40.99}{\percent} to \SI{83.18}{\percent}.

\subsubsection{Summary} To summarize, \cvfrhq is advantageous in scenarios where bandwidth conservation is a primary concern, such as in regions with limited network capacity or for streaming providers aiming to reduce data delivery costs. Despite higher encoding energy consumption, \cvfrhq delivers superior video quality due to its efficient utilization of available bits, contributing to a more satisfying streaming experience. Meanwhile, \cvfreco prioritizes energy efficiency, lowering computational resource consumption, which is suitable for reducing energy consumption in data centers. \cvfreco can still offer bandwidth optimization, albeit to a lesser extent, compared to \cvfrhq. The choice of the encoding scheme should align with the overarching objectives and constraints of the streaming service provider and its target audience.

%% file: sec_conclusions.tex
\section{Conclusions}
\label{sec:conclusion_future_dir}
This paper proposes a content-adaptive variable framerate encoding scheme (\cvfr) for adaptive live-streaming applications. \cvfr includes a random forest-based model, which predicts optimized framerate and encoding preset for each bitrate ladder representation of a given video segment based on its spatiotemporal characteristics. Furthermore, a JND-based representation elimination algorithm is proposed to minimize the perceptual redundancy of the representations. Two variations of \cvfr: \cvfreco and \cvfrhq are presented, where the former predicts an optimized framerate for each representation using the fastest preset. The experimental results show that, on average, \cvfreco yields bitrate savings of \SI{17.91}{\percent} and \SI{13.54}{\percent} to maintain the same PSNR and VMAF, respectively, compared to the fixed framerate-fastest preset CBR encoding of the reference HLS bitrate ladder using x264 encoder. This is accompanied by a cumulative decrease of \SI{48.64}{\percent} in encoding energy needed for various representations and \SI{76.24}{\percent} in storage energy, considering a JND of six VMAF points. However, \cvfrhq predicts the optimized framerate-preset pair for each representation, which yields the highest possible compression efficiency within the low latency encoding speed threshold. The experimental results show that, on average, \cvfrhq yields bitrate savings of \SI{54.25}{\percent} and \SI{39.63}{\percent} to maintain the same PSNR and VMAF, respectively, compared to the fixed framerate-fastest preset CBR encoding of the reference HLS bitrate ladder. Although the encoding energy consumption increases to \SI{282.94}{\percent}, \cvfrhq yields an \SI{83.18}{\percent} decrease in storage energy, respectively, considering a JND of six VMAF points.

In the future, \cvfr can be extended to high-framerate video streaming applications, where the original video framerates are up to 120\,fps. \cvfreco can be extended for VoD applications using the slowest encoding preset and sophisticated temporal filtering and frame interpolation methods, as the low-latency constraint can be compromised.